\documentclass[aps,prr,superscriptaddress,amsfonts,amsmath,amssymb,showpacs,floatfix,reprint,longbibliography]{revtex4-2}
\usepackage{url}
\usepackage{bm}
\usepackage{graphicx}
\usepackage{amstext}
\usepackage{amsbsy}
\usepackage{color}
\usepackage[colorlinks=true, urlcolor=blue, linkcolor=blue, citecolor=blue, pdftex]{hyperref}
\usepackage{multirow}
\usepackage{float}
\usepackage{times}
\usepackage[utf8]{inputenc}

\newcommand{\ket}[1]{{\left | #1 \right\rangle}}
\newcommand{\bra}[1]{{\left \langle #1 \right|}}

\usepackage{tikz}

\newcommand{\vertIa}{\tikz[baseline={([yshift=-.5ex]current bounding box.center)},x=0.5em,y=0.5em]{
  \draw[very thick] (-1,0) -- (0,0) -- (1,0); \draw[very thin] (0,-1) -- (0,0) -- (0,1)}}
\newcommand{\vertIb}{\tikz[baseline={([yshift=-.5ex]current bounding box.center)},x=0.5em,y=0.5em]{
    \draw[very thick] (0,-1) -- (0,0) -- (0,1); \draw[very thin] (-1,0) -- (0,0) -- (1,0)}}
\newcommand{\vertIIa}{\tikz[baseline={([yshift=-.5ex]current bounding box.center)},x=0.5em,y=0.5em]{
    \draw[very thick] (-1,0) -- (0,0) -- (0,1); \draw[very thin] (1,0) -- (0,0) -- (0,-1)}}
\newcommand{\vertIIb}{\tikz[baseline={([yshift=-.5ex]current bounding box.center)},x=0.5em,y=0.5em]{
    \draw[very thick] (0,1) -- (0,0) -- (1,0); \draw[very thin] (0,-1) -- (0,0) -- (-1,0)}}
\newcommand{\vertIIc}{\tikz[baseline={([yshift=-.5ex]current bounding box.center)},x=0.5em,y=0.5em]{
    \draw[very thick] (1,0) -- (0,0) -- (0,-1); \draw[very thin] (-1,0) -- (0,0) -- (0,1)}}
\newcommand{\vertIId}{\tikz[baseline={([yshift=-.5ex]current bounding box.center)},x=0.5em,y=0.5em]{
    \draw[very thick] (0,-1) -- (0,0) -- (-1,0); \draw[very thin] (0,1) -- (0,0) -- (1,0)}}
    
\newcommand{\vertHplaq}{\tikz[baseline={([yshift=-.5ex]current bounding box.center)},x=0.45em,y=0.45em]{
  \draw[very thick] (0,0) -- (2,0);  \draw[very thick] (0,2) -- (2,2); \draw[very thin] (0,0) -- (0,2); \draw[very thin] (2,0) -- (2,2)}}

\newcommand{\vertVplaq}{\tikz[baseline={([yshift=-.5ex]current bounding box.center)},x=0.45em,y=0.45em]{
  \draw[very thin] (0,0) -- (2,0);  \draw[very thin] (0,2) -- (2,2); \draw[very thick] (0,0) -- (0,2); \draw[very thick] (2,0) -- (2,2)}}

\newcommand{\dimerH}{\tikz[baseline={([yshift=-.5ex]current bounding box.center)},x=0.45em,y=0.45em]{
  \draw[very thick] (0,0) -- (2,0); }}

\newcommand{\dimerV}{\tikz[baseline={([yshift=-.5ex]current bounding box.center)},x=0.45em,y=0.45em]{
  \draw[very thick] (0,0) -- (0,2); }}

\begin{document}

\title{
Crystalline phases and devil's staircase in qubit spin ice 
}

\author{M\'ark Kond\'akor}
\affiliation{Institute of Physics,
Budapest University of Technology and Economics, M\"uegyetem rkp. 3., H-1111 Budapest, Hungary}
\affiliation{Institute for Solid State Physics and Optics, Wigner Research Centre for Physics, H-1525 Budapest, P.O.B. 49, Hungary}
\affiliation{Max-Planck-Institut f\"ur Festk\"orperforschung,
Heisenbergstr. 1, 70569 Stuttgart, Germany}
\author{Karlo Penc}
\affiliation{Institute for Solid State Physics and Optics, Wigner Research Centre for Physics, H-1525 Budapest, P.O.B. 49, Hungary}

\date{\today}

\begin{abstract}
Motivated by the recent realization of an artificial quantum spin ice in an array of superconducting qubits with tunable parameters [King {\it et al.}, Science 373, 576 (2021)], we scrutinize a quantum six vertex model on the square lattice that distinguishes type-I and type-II vertices. We map the zero-temperature phase diagram using numerical (exact diagonalization) and analytical (perturbation expansion, Gerschgorin theorem) methods. Following a symmetry classification, we identify three crystalline phases alongside a subextensive manifold of isolated configurations. Monte Carlo simulations at the multicritical Rokhsar-Kivelson point provide evidence for a quantum phase exhibiting a cascade of transitions with increasing flux. By comparing structure factors, we find evidence for the emergence of the fully flippable and plaquette phases in the artificial quantum spin ice.
\end{abstract}


\maketitle

\section{Introduction}
\label{sec:intro}

\begin{figure}[b]
\includegraphics[width=.7\columnwidth]{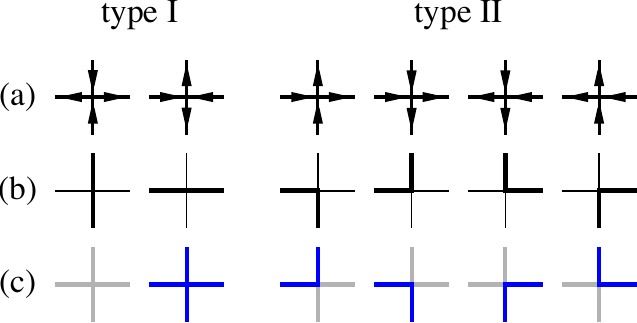}
\caption{
(a) The two-in, two-out configurations in the six-vertex model. While all of them are equivalent in three dimensions, they become distinguishable for the two-dimensional ice and split into type-I and type-II vertices.
(b) In the fully packed loop representation, the thick bonds are occupied (i.e., part of a loop), whereas the thin bonds are empty. The figure shows the correspondence between the arrows and the occupied or vacant bonds for "even" sites, while the occupancies are reversed for "odd" sites. The "even/odd" refers to the parity of the sum $i_x+i_y$ of a site's coordinates $(i_x,i_y)$ in the lattice.
(c) In the alternative Baxter representation, the bonds are colored blue if the arrows on the bond are reversed compared to the reference vertex in the first column.
\label{fig:vertex_types}
}
\end{figure}

Water ice is a beautiful example of how an everyday material can inspire and advance many areas of physics, from classical statistical physics to qubit systems.
Oxygen ions in ice form a four-fold coordinated lattice. The protons are located on the links connecting the nearest neighbor oxygen ions: two protons bond to the oxygen covalently, and two with hydrogen bonds. It is the ice rule identified by Bernal and Fowler in 1933~\cite{BernalFowler1933}. 
Because the lengths of the covalent and hydrogen bonds are different, arrows can be used to show the position of the protons about the center of the link.
It leads to the ''two-in, two-out" formulation of the ice rule, which can be considered a local divergence-free condition.
The ice rule allows for six different proton configurations around an oxygen ion.
Because these six configurations correspond to six different vertices in the arrow representation [see Fig.~\ref{fig:vertex_types}(a)], the fundamental model of ice is known as the six-vertex model (6VM).
Remarkably, the number of states satisfying the ice rule increases exponentially with the size of the system, forming a manifold.
Pauling estimated the degeneracy of the manifold as $W_{\text{ice}}=(3/2)^{N} = 1.5^N$, where $N$ is the number of oxygen ions, resulting in a finite residual entropy~\cite{Pauling1935}. 
Lieb solved the two-dimensional six-vertex model on the square lattice exactly and got $W_{\text{2D}}= (4/3)^{3N/2} \approx 1.5396^N$ \cite{Lieb_entropy_PhysRevLett.18.692}, which is very close to Pauling's estimate.
Furthermore, Baxter noted that the correlations decay algebraically \cite{Baxter}.

The six-vertex model successfully describes physical systems in which divergence-free conditions arise. For instance, Anderson proposed that the frustration in magnetite leads to charge disproportionation, where configurations following the Bernal-Fowler rule minimize the Coulomb energy \cite{Anderson1956}.  
Another example is the ``square ice'' substance KH$_2$PO$_4$ (KDP), a quasi 2D material in which the vertex configurations do not have equal energies \cite{Slater_KDP_1941, Schmidt1987}. 
In particular, the discovery of spin ice materials brought the field to flourish \cite{Harris_PhysRevLett.79.2554}. 
In spin ice, the Ising-like magnetic moments of rare earth ions form the
highly frustrated pyrochlore lattice of corner-sharing tetrahedra with spins at the corners. Arrows representing these spins realize the low energy two-in, two-out configurations~\cite{Ramirez1999,*Bramwell_Gingras_Science_2001}. 
Since then, spin ice physics was also accomplished in fabricated arrays of nanomagnets \cite{Artificial_spin_Ice_Wang_Nature_2006, Moeller_PRL_2006,Nisoli_2013RvMP...85.1473N, Artificia_spin_ice_review_Nat_Rev_2020}.


The six-vertex model is a convenient starting point for realizing topological defects due to its correlated ground state manifold. 
Flipping an arrow creates a ``three-in, one-out" and a ``one-in, three-out" vertex. These vertices are not part of the six-vertex model; the underlying model (e.g., Ising model) determines their dynamics. 
They correspond to fractional charges in the model for magnetite \cite{Fulde_Penc_Shannon_2002,*PhysRevB.70.245113,*PhysRevLett.97.170407} and to the emergence of magnetic monopoles in the spin ice systems \cite{Ryzhkin_2005,*Castelnovo2008}, experimentally confirmed in Ref.~\cite{doi:10.1126/science.1177582,*doi:10.1126/science.1178868}.


\begin{figure}[tb]
\includegraphics[width=.75\columnwidth]{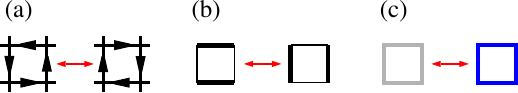}
\caption{
(a) The off-diagonal term $\mid \circlearrowleft \rangle\langle \circlearrowright \mid \!+\!
\mid \circlearrowright \rangle\langle \circlearrowleft
\mid$ reverses the direction of arrows along a directed loop around a square plaquette. 
(b) In the fully packed loop representation, the off-diagonal term
acts on configurations where a pair of opposite bonds is empty and the other is occupied. 
(c) In the alternative Baxter representation, the four gray colors become blue and vice versa.\label{fig:PlaquetteFlipping}
}
\end{figure}

 The quantum six-vertex model emerges by allowing tunneling between configurations that obey the ice rule by adding an off-diagonal term to the classical Hamiltonian. 
This term reverses arrows arranged tail-to-nose around an elementary square plaquette, as shown in Fig.~\ref{fig:PlaquetteFlipping}a, though longer loops are also allowed.
Chakravarty used such an expression to describe $d$-wave superconductors in Ref.~\onlinecite{Chakravarty_PRB_2002}. 

The quantum six-vertex model appears in the perturbative expansion of the $S=1/2$ Heisenberg model on lattices of corner-shared tetrahedra in the limit of large easy-axis exchange anisotropy.
For example, on the 2-dimensional checkerboard lattice, the quantum term gives rise to a gapped phase, where arrows on alternating square plaquettes resonate \cite{Shannon2004, Syljuasen_PRL_2006}.
On the 3-dimensional pyrochlore lattice, Hermele {\it et al.} argued that the effective theory is a Maxwellian $U(1)$ action with gapless ``photon''-like excitations \cite{Hermele2004}, confirmed numerically in Refs.~\onlinecite{Banerjee2008, Shannon2012, Pace_PRL_2021}. 
This situation may arise in certain spin-ice materials:  Tb$_2$Ti$_2$O$_7$~\cite{Molavian_PhysRevLett.98.157204_2007,*molavian2009effective}, 
 Pr$_2$Sn$_2$O$_7$ and Pr$_2$Zr$_2$O$_7$~\cite{Onoda_PhysRevLett.105.047201,*SungBin_PhysRevB.86.104412_2012,*Kimura:2013aa},
 and Yb$_2$Ti$_2$O$_7$~\cite{Ross_PhysRevX.1.021002_2011} are all suitable candidates (see  Ref.~\onlinecite{Gingras_McClarty_2014RPPh...77e6501G} for a review about the quantum spin ice).
The stability of the $U(1)$ spin liquid against ordered phases in quantum spin ice was considered in  Refs.~\onlinecite{Lucile_PhysRevLett.108.037202, Benton_PhysRevB.86.075154_2012}, together with experimental signatures. 

Other examples include the isotropic $S=1/2$ Heisenberg model with four-site ring exchange on checkerboard and pyrochlore lattices, with the constraint of exactly one singlet bond on each tetrahedron~\cite{Nussinov_PRB_2007}. 
By extending the fundamental model, Ref. \cite{Nagaosa_PhysRevLett.112.247602} investigated the quantum effects in the ''square ice'' KH$_2$PO$_4$. 
Ref.~\cite{Roscilde_PhysRevLett.113.027204} studied finite temperature effects in quantum square ice. 
More recently, the dynamical properties of the model came under scrutiny: it exhibits dynamical quantum phase transitions \cite{PhysRevLett.122.250401} and quantum many-body scars~\cite{Banerjee2021,*10.21468/SciPostPhys.12.5.148}.
 
The connection between gauge theories and the quantum six-vertex model, initially discussed in Ref.~\cite{Hermele2004}, was further explored in Ref.~\cite{Castro_Neto_PhysRevB.74.024302_2006}, where the two-dimensional quantum six-vertex model was found to be a confining lattice gauge model. 
The model is also known as the (2+1)-dimensional $U(1)$ quantum link model~\cite{ORLAND1990647,*CHANDRASEKHARAN1997455}. Refs.~\onlinecite{Banerjee_2013, Tschirsich_MPS_SciPost_2019} considered the model from a gauge field theory point of view. 
This relationship inspired the concept of engineering arrays of Rydberg atoms as simulators of $U(1)$ lattice gauge theories in various geometries~\cite{PhysRevX.4.041037, PhysRevX.10.021057, Ran2022}.

The motivation for our research comes from the recent implementation of the quantum six-vertex model in a quantum annealing system by King et al. \cite{King2021}. 
Their setup consisted of superconducting flux qubits arranged in an array that physically realized the transverse-field Ising model on the checkerboard lattice. 
Four ferromagnetically coupled qubits formed a single logical spin, representing an Ising spin. 
Antiferromagnetic two-body couplers between qubits belonging to adjacent logical spins provided a tunable antiferromagnetic interaction between the Ising spins. 
They implemented two inequivalent couplers that enabled tuning the parameters of the Ising model into the range described by the six-vertex model and lifting the degeneracy between type-I and type-II vertices (see Fig.~\ref{fig:vertex_types}a). 
Quantum fluctuations induced by the transverse field led to tunneling between the six-vertex configurations. 
Thus, the minimal model of their setup involves the tunneling term 
\begin{subequations}
\begin{equation}
  \mathcal{H}_{t} = -t \sum_{\rm plaq.}
 \bigl(
\mid \circlearrowleft \rangle\langle \circlearrowright \mid \!+\!
\mid \circlearrowright \rangle\langle \circlearrowleft
\mid \bigr),
  \label{eq:Ht}
\end{equation}
where the sum is over the elementary square plaquettes of the lattice, and $\mid \circlearrowright \rangle$ and $\mid \circlearrowleft \rangle$ denote square plaquettes with the clockwise and anticlockwise orientation shown in  Fig.~\ref{fig:PlaquetteFlipping}(a), and a chemical potential 
\begin{equation}
\mathcal{H}_{\text{II}}  =  \mu \hat{N}_{\rm II}
  \label{eq:HII}
\end{equation}
to distinguish the two types of vertices (the $\hat{N}_{\text{II}}$ operator counts the number of the type-II vertices). 
It is also convenient to introduce the 
\begin{equation}
 \mathcal{H}_{V} = V \sum_{\rm plaq.}
 \bigl(
 \mid \circlearrowleft \rangle\langle \circlearrowleft \mid \!+\!
 \mid \circlearrowright \rangle\langle \circlearrowright \mid \bigr) = V \hat{N}_{V},
   \label{eq:HV}
\end{equation}
\end{subequations}
term, where $\hat{N}_V$ counts the number of flippable plaquettes.
The full model we will consider in this paper is then
\begin{equation}
  \mathcal{H} =
  \mathcal{H}_{t} + \mathcal{H}_{V} + \mathcal{H}_{\text{II}}.
  \label{eq:Q6VMani}
\end{equation} 
Let us briefly review the known limiting cases of the Hamiltonian. 

$\mathcal{H}_{\text{II}}$ is the Hamiltonian of the Rys $F$ model \cite{Rys1963}. It is a well-known problem in statistical physics \cite{Baxter} and exactly solvable by Bethe Ansatz \cite{Lieb1967, Sutherland1967_PhysRevLett.19.103}. 
It exhibits two phases at zero temperature.
For $\mu>0$, the two-fold degenerate ground state consists of alternating 
type-I vertices -- this is the antiferroelectric phase. 
If $\mu<0$, configurations with only type-II vertices span the disordered phase's highly degenerate ground state manifold.

\begin{figure}[bt]
\includegraphics[width=.75\columnwidth]{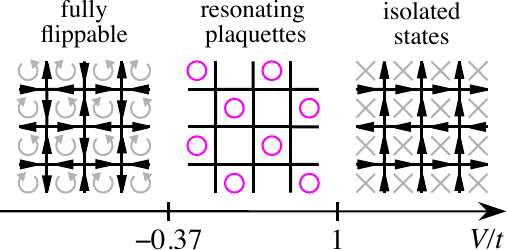}
\caption{
 The phase diagram of the quantum six-vertex model, defined by Hamiltonian~(\ref{eq:Q6VMoriginal}), consists of three phases. 
 For $V \lesssim -0.37t$, the twofold degenerate fully flippable phase arises, where the number of flippable plaquettes is maximal. 
 Every second plaquette (denoted by magenta circles) resonates when $-0.37 t \lesssim V \leqslant t$. 
 The approximate wave function is a direct product of the 
$| \bigcirc \rangle = \frac{1}{\sqrt{2}} (\mid \circlearrowright \rangle + \mid \circlearrowleft \rangle) $ on the resonating plaquettes [$| \bigcirc \rangle = \frac{1}{\sqrt{2}} (| \protect\vertHplaq \rangle + | \protect\vertVplaq \rangle) $ in the loop representation]. 
The Rokshar-Kivelson point at $V=t$ is quantum critical, and for $t<V$, the isolated manifold appears with configurations having no flippable plaquettes (denoted by crosses).
\label{fig:Q6VMoriginal}
}
\end{figure}

Following the footsteps of Rokshar and Kivelson \cite{RokhsarKivelson1988}, Shannon {\it et al.} went beyond the pure quantum six vertex model $\mathcal{H}_{t}$ of Chakravarty~\cite{Chakravarty_PRB_2002} and introduced the 
\begin{equation}
  \mathcal{H}_{0} =
  \mathcal{H}_{t} + \mathcal{H}_{V}  \label{eq:Q6VMoriginal}
\end{equation}
Hamiltonian \cite{Shannon2004}. In the fully-packed loop representation, the Hamiltonian (\ref{eq:Q6VMoriginal}) has precisely the same form as the quantum-dimer model of Rokshar and Kivelson, except for the Hilbert space: it acts on dimers in one case and loops in the other. 
Just like in the quantum-dimer model, the exact ground state of the model is an equal-weight superposition of all connected states when $V=t$. 
This is a quantum critical point with algebraically decaying correlations. It separates the subextensively degenerate ground state manifold of isolated (also called disconnected) states from the resonating plaquette phase (see Fig.~\ref{fig:Q6VMoriginal} for a sketch of the phase diagram).
Configurations in the isolated manifold consist of type-II vertices only and have no flippable plaquettes; thus, $\mathcal{H}_{t}$ annihilates them. 
The resonating plaquette phase is similar to the one in the quantum dimer model~\cite{Leung_QDM_PhysRevB.54.12938}, but every alternating square plaquette resonates, so it is two-fold degenerate only.
Exact diagonalization studies estimated the lower boundary of the plaquette phase as $V/t \approx -0.3727$~\cite{Shannon2004} and $V/t = -0.359(5)$~\cite{Banerjee_2013}, a gauge-invariant matrix product states calculation located the transition point at $V/t = -0.37(3)$~\cite{Tschirsich_MPS_SciPost_2019}, and quantum Monte Carlo at $V/t = -0.35(3)$~\cite{Ran2022}. 
Below this boundary, flippable plaquettes minimize the energy but without resonance. One can think of this ``fully flippable phase" as the antiferroelectric phase of the Rys $F$ model dressed with quantum fluctuations. This phase corresponds to the N\'eel phase in the $XXZ$ model on the checkerboard lattice~\cite{Shannon2004}. 

The present study aims to extend the phase diagram introduced above and shown in Fig.~\ref{fig:Q6VMoriginal} by including the term with the chemical potential for the type-II vertices, Eq.~(\ref{eq:HII}), noting that for $V=0$ and $\mu>0$, the plaquette phase is known to persist up to $\mu/t = 0.288$ \cite{Syljuasen_PRL_2006}. We will derive the phase diagram of the quantum spin ice Hamiltonian (\ref{eq:Q6VMani}) in the complete $V$--$\mu$ plane and the structure factor at zero temperature. It will allow us to get an insight into the results of the qubit-engineered quantum spin ice of King {\it et al.} \cite{King2021}.  


The paper is organized as follows. 
We describe the classical six-vertex configurations in various representations and the flux sectors in finite-size clusters with periodic boundary conditions in Sec.~\ref{sec:classical}. In  Sec.~\ref{sec:CPD}, we construct the $t=0$ classical phase diagram of the model.
In Sec.~\ref{sec:Symmetries}, we systematically classify the symmetries of the model, construct order parameters, and write the Landau free energy for phases with zero topological flux.
Sec.~\ref{sec:isolated} discusses the properties of isolated states. 
In Sec.~\ref{sec:ED}, we present numerical results (exact diagonalization) to reveal the ground state phase diagram of the quantum model.
As an independent check, in Sec.~\ref{sec:perturb}, we use perturbation theory to calculate corrections to the ground state energies of the classical phases and deduce some of the phase boundaries. 
In Sec.~\ref{sec:rainbow}, we sample the wave function at the Rokhsar-Kivelson point with a Monte Carlo method and explore the phases emanating from this quantum critical point. We also characterize the emergent quantum electrodynamics.
Structure factors in different phases are evaluated in Sec.~\ref{sec:structure_factors} and compared with the ones observed in qubit quantum spin ice.
We conclude with a summary of results in Sec.~\ref{sec:summary}. 
Finally, appendices \ref{sec:Appendix_flux_sectors}-\ref{sec:SqPlaq} contain some details of our calculations.

\section{The six-vertex configurations}
\label{sec:classical}

\begin{figure}[tb]
\includegraphics[width=.95\columnwidth]{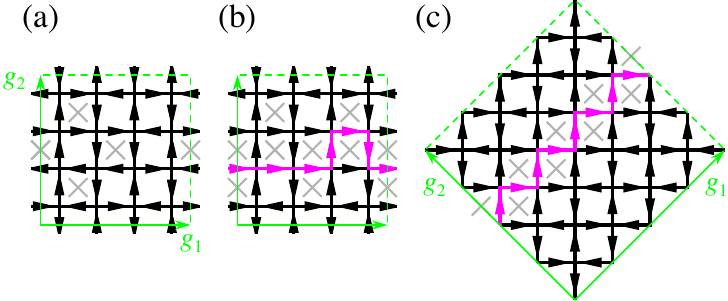}
\caption{\label{fig:Clusters}
(a) The $N=16$ square-shaped cluster with 16 sites and periodic boundary conditions is defined by the lattice vectors $g_1=(4,0)$ and $g_2 =(0,4)$. The shown ice-rule obeying configuration, with 12 type-I and four type-II vertices, is in the $(0,0)$ flux sector. All the plaquettes are flippable except for the four denoted by crosses. (b) Reversing the arrows along the magenta loop crossing the boundaries, we get a configuration in the $\mathbf{m}=(2,0)$ flux sector. (c) The $N=32$ site cluster defined by $g_1=(4,4)$ and $g_2 =(-4,4)$ from the $N=2L^2$ family of cluster. The presented configuration is in the lowest non-zero $\mathbf{m}=(2,2)$ flux sector. It originates from a periodic configuration of arrows where all the plaquettes are flippable (we call it a fully flippable configuration later on) by reversing the arrows along the magenta path.}
\end{figure}

\subsection{Finite clusters}
\label{subsec:FiniteClusters}

We study the six-vertex model on finite clusters with periodic boundary conditions on the square lattice. Their size $N$ and geometrical symmetries characterize these clusters.  We focus on two families having the full $\mathsf{D_4}$ point group symmetry of the lattice. We refer to the ones generated by the $\mathbf{g}_1 = (L,0)$ and $\mathbf{g}_2 = (0,L)$ lattice vectors as the $N=L^2$ family, see Fig.~\ref{fig:Clusters}(a). The $\mathbf{g}_1 = (L,L)$ and $\mathbf{g}_2 = (-L,L)$ lattice vectors define the $N=2L^2$ family, shown in Fig.~\ref{fig:Clusters}(c). The periodicity of the ground states, as we will see later on, requires even values for $L$. We consider lattice sites translated by an integer multiple of the $\mathbf{g}_1$ and $\mathbf{g}_2$ identical, $\mathbf{r} + n_1 \mathbf{g}_1 + n_2 \mathbf{g}_2 \cong \mathbf{r} $, where $n_1,n_2 \in \mathbb{Z}$.

\subsection{Representations of the 6-vertex configurations}
\label{sec:app_representations}

\begin{figure}[bt!]
        \centering
        \includegraphics[width = 0.8\columnwidth]{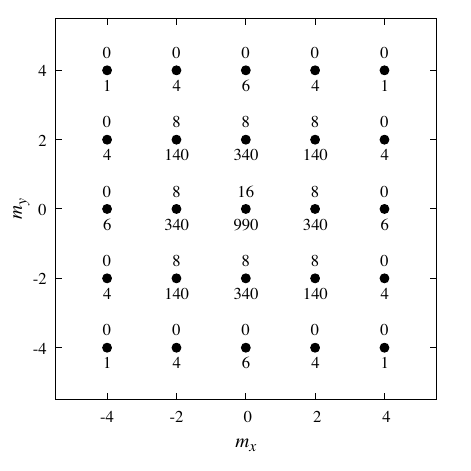}
    \caption{The degeneracy of states (below) and maximal value of $N_V$ (above) in the different flux sectors $(m_x,m_y)$ of the 16-site cluster with periodic boundary conditions.
 \label{fig:flux_16}}
\end{figure}

The classical 6-vertex model has a long history and applies to many systems, each prompting a convenient representation. Below we review some of them. 

\paragraph{Arrow representation:} This is the original representation of water ice. A configuration is represented as a directed graph, with arrows showing the direction of the edges (bonds). To satisfy the ice rule, every vertex has two inward and two outward pointing arrows, demonstrated in Fig.~\ref{fig:vertex_types}(a). A plaquette is flippable if the arrows around the elementary square point clockwise or counterclockwise. The flip itself corresponds to changing the directions of the arrows around a plaquette; see Fig.~\ref{fig:PlaquetteFlipping}(a). This representation is meaningful for calculating neutron scattering cross section detailed in Sec.~\ref{sec:structure_factors}. 

\paragraph{Fully packed loop representation:} In Ref.~\cite{Anderson1956}, Anderson described magnetite as a charge-frustrated material using the Ising model. The charge frustrated Fe$^{+2.5}$ build a pyrochlore lattice consisting of corner-shared tetrahedra. The minimal Coulomb energy corresponds to two $2.5+\delta q$ and two $2.5-\delta q$ charged ions on each tetrahedron. These are represented as occupied and empty bonds, shown in Fig.~\ref{fig:vertex_types}(b) for the two-dimensional model on the checkerboard lattice. Identically charged bonds form closed loops in a finite system. 
A plaquette is flippable if, as we go around it, oppositely charged bonds meet at each vertex; see Fig.~\ref{fig:PlaquetteFlipping}(b). 
The bonds exchange their charges by a plaquette flip, just like in the quantum dimer model \cite{Kivelson_QDM_1987} describing short-range resonating valence bonds.  

We measure the occupancy of a bond by $n_{\mathbf{r}}=\pm 1$, where ${\mathbf{r}}$ is the coordinate of the center of the bond.
Assuming that the horizontal bonds are along the $x$ and the vertical along the $y$ direction, the following relations hold between the arrow and fully packed loop representations:
\begin{subequations}
\label{eq:FPL2Arrows}
\begin{align}
   \mathbf{M}_{(i_x+1/2,i_y)} &= (-1)^{(i_x+i_y)} n_{(i_x+1/2,i_y)} 
   \begin{pmatrix}
    1 \\
    0 
  \end{pmatrix}, \\
   \mathbf{M}_{(i_x,i_y+1/2)} &= (-1)^{(i_x+i_y)} n_{(i_x,i_y+1/2)} 
   \begin{pmatrix}
    0 \\
    1 
  \end{pmatrix}.
\end{align}
\end{subequations}
The integer-valued $(i_x,i_y)$ are the coordinates of the vertices, and the bond lengths are set to 1.

\paragraph{Baxter- and alternative Baxter-representation:} In his textbook \cite{Baxter}, Baxter chose an isolated configuration formed by identically oriented horizontal and vertical bonds as a reference configuration, shown in Fig.~\ref{fig:iso_states}(a). Then he highlighted all the bonds in a configuration that pointed in the opposite direction compared to the reference. Here we use the same principle but choose one of the fully-flippable configurations as the reference (therefore, we call it the alternative Baxter representation); see Fig.~\ref{fig:vertex_types}(c). For instance, a vertex is type-II if two highlighted bonds meet there, and a plaquette is non-flippable if it has both highlighted and non-highlighted bonds [Fig.~\ref{fig:PlaquetteFlipping}(c)]. This representation helps us to identify the mathematical structure of the configuration space. 

\paragraph{Faraday loop representation:} Type-II vertices can be associated with local dipole moments. Drawing these dipole moments as arrows, they form closed loops in ice rule obeying configurations~\cite{Nisoli_2020}. They help study the thermodynamic properties of ice systems and provide a way to approach magnetic monopoles.

\paragraph{Height representation:} The local divergence-free constraint at the vertices enables us to transcribe an arrow configuration to integers on the plaquettes. Since we are not using it in our work, we only refer to Refs.~\cite{Henley:1997aa} for details.

All of the representations above constitute a basis where both the $\hat N_V$ and $\hat N_{\text{II}}$ operators are diagonal, and the quantum flipping term $\mathcal{H}_t$ is strictly off-diagonal.
We refer the reader to Ref.~\cite{zinnjustin2009sixvertex} for a comprehensive account of the various six-vertex model representations.

\subsection{Flux sectors}

In a cluster with periodic boundary conditions, for each 6-vertex configuration, we can count the net flux of arrows through any given vertical ($m_x$) or horizontal ($m_y$) cut. Since the local flips do not change the net flux, the vector $\mathbf{m} = (m_x,m_y)$ defines a set of winding numbers that the Hamiltonian conserves. States having the same index pair $ (m_x,m_y)$ form a flux sector. 
We can generate configurations in different flux sectors by flipping arrows on a directed loop crossing the cluster's boundaries, as illustrated in Fig.~\ref{fig:Clusters}(b). In the $N=2L^2$ clusters, the horizontal and vertical cuts are the diagonals of the rotated square, and the minimal nonzero flux sector is the $\mathbf{m} = (2,2)$ shown in Fig.~\ref{fig:Clusters}(c). 
Let us note that the $\mathbf{m}$ defining the flux sectors is proportional to the total magnetization in the Faraday loop description \cite{Nisoli_2020}, and the quantum term mixes all the configurations having the same total magnetization (except the isolated states).

Besides the geometric symmetries described by the point and translation group of the lattice or cluster, there is an internal symmetry, the charge conjugation $\mathcal{C}$~\cite{Banerjee_2013}. It reverses the occupation of the bonds in the fully packed loop representation and commutes with the Hamiltonian, $[\mathcal{C},\mathcal{H}]=0$.  In the arrow representation, it reverses the direction of all the arrows. As a consequence, the flux sector of a configuration changes signs under charge conjugation,  $\mathcal{C} \mathbf{m} = -\mathbf{m}$. In the alternative Baxter representation,  $\mathcal{C}$ changes highlighted edges into unlighted ones and vice versa.

We mostly use numerical means to calculate the ground state properties. To this end, we shall generate all the possible ice-rule configurations in a given cluster. Based on empirical findings on small clusters, we assume that the flux sectors are ergodic if their classical states have at least one flippable plaquette (ergodicity has been proven for the $\mathbf{m}=(0,0)$ flux sector in Ref.~\onlinecite{Hermele2004}). 
Therefore, it is enough to find a configuration from a flux sector, since
applying local flips will generate all the configurations within the sector. 
In the case of the $N=L^2$ clusters, one can find a systematic way to construct initial configurations using only type-II vertices. In these configurations, the directions of the arrows along a horizontal or vertical line are all the same (but the directions may differ from line to line). Turning all the arrows on a line changes the flux sector by one unit, allowing access to the desired flux sector.

In Fig.~\ref{fig:flux_16}, we present the number of configurations in each flux sector for the $N=16$ site cluster. The dimension of the Hilbert space in the $\mathbf{m} = (0,0)$ flux sector is a modest 990. Data for larger site clusters are presented in Fig.~\ref{fig:topsec_32_36} in Appendix~\ref{sec:Appendix_flux_sectors}. We just note that the dimensions of the $(0,0)$ flux sectors are $962\,734$ for $N=32$ and $5\,482\,716$ for $N=36$, these are easy to diagonalize by the L\'anczos method.

\begin{figure}[bt]
\includegraphics[width=.8\columnwidth]{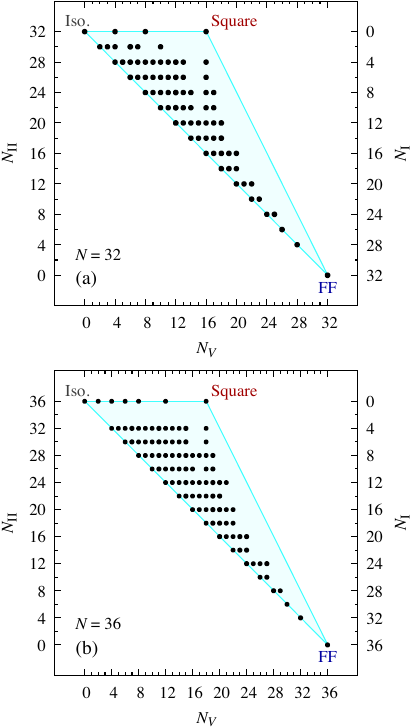}
\caption{The map shows the total number of flippable plaquettes ($N_V$)  as well as the number of type-I and type-II vertices  ($N_{\text{I}}$ and $N_{\text{II}}$, where $N_{\text{I}}+N_{\text{II}} = N$) in the classical basis of the six-vertex configurations. The convex hull is a triangle, and the two fully flippable (FF) configurations shown in Fig.~\ref{fig:FFs}, the four square (Sq) configurations of Fig.~\ref{fig:Squares}, and the isolated manifold, shown Fig.~\ref{fig:iso_states}(a) and (c), are located at the corners.
\label{fig:nVnII_map}
}
\end{figure}

Plotting the possible $N_V$ and $N_{\text{II}}$ values of the configurations, we find these values are not independent. Fig.~\ref{fig:nVnII_map} shows a map for the $N=32$ and $N=36$ site clusters. The ice rule and the periodic boundary conditions constrain the number of allowed type-II vertices and flippable plaquettes to a triangle in the $N_{\text{II}}$--$N_V$ plane. In Appendix~\ref{sec:inequalities} we derive the inequalities
\begin{subequations}
   \label{eq:GeomConstrains}
   \begin{align}
        & N_{\rm II} \leq N \;, \label{eq:GeomConstrainsA}\\
        & 2 N_{V} + N_{\rm II} \leq 2 N \;, \label{eq:GeomConstrainsB}\\
        & N \leq N_{V} + N_{\rm II} \;, \label{eq:GeomConstrainsC}
    \end{align}
\end{subequations}
which determine the triangle boundaries for a cluster with $N$ vertices (i.e., $N$ sites).

\section{Classical phase diagram}
\label{sec:CPD}

\begin{figure}[bt!]
\includegraphics[width=.8\columnwidth]{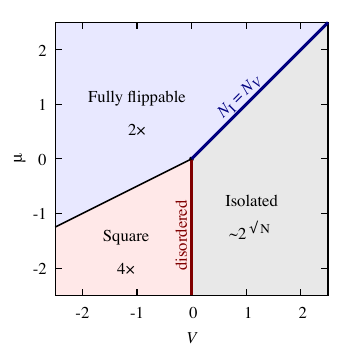}
\caption{The classical phase diagram in the parameter space of the $V$ and $\mu$, the chemical potential of the type-II vertices. The isolated and square phases consist only of type-II vertices; they become favorable when $\mu$ is negative and $V$ is positive. The fully flippable phase consists of type-I vertices, and all plaquettes are flippable, gaining energy when $V$ is negative. There are two fully flippable states (shown in Fig.~\ref{fig:FFs}) and four square states (see Fig.~\ref{fig:Squares}). The degeneracy of the isolated phase is subextensive and increases exponentially with the linear size of the cluster. The boundary between the square phase and isolated manifold (thick red line) hosts the disordered manifold of the Rys $F$ model.
\label{fig:CPD}}
\end{figure}

Below we derive the phase diagram in the classical limit of the Q6VM where $t$ vanishes. The Hamiltonian is diagonal in the basis of both the six-vertex and fully packed loop configurations shown in Fig.~\ref{fig:vertex_types}. The energy of a configuration depends only on the number of type-II vertices $N_{\text{II}}$ and  flippable plaquettes $N_V$ as
\begin{equation}
\mathcal{H}^{\text{cl}}  = V \hat{N}_{V}  + \mu \hat{N}_{\text{II}} \,.
\label{eq:ClassicalEnergy}
\end{equation}
Since the energy in Eq.~(\ref{eq:ClassicalEnergy}) is linear in both $N_{\text{II}}$ and $N_V$, three phases emerge in the minimization procedure corresponding to the three corners of the triangle. Fig.~\ref{fig:CPD} shows the classical phase diagram.

The first phase consists of isolated configurations having no flippable plaquettes and only type-II vertices so that $(N_V, N_{\text{II}}) = (0, N)$. The energy is then
\begin{equation}
  E_{\text{Iso}} = \mu N \,.
  \label{eq:Eiso}
\end{equation}
%
We devote section~\ref{sec:isolated} to the properties of the isolated manifold. 

\begin{figure}
\includegraphics[width=.45\columnwidth]{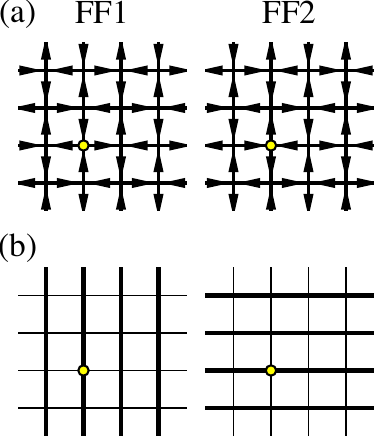}
\caption{The two fully flippable configurations in (a) 6-vertex and (b) fully packed loop representation. The open circles serve as anchor points.
\label{fig:FFs}
}
\end{figure}

The second one is the fully flippable phase (FF). It maximizes the number of the flippable plaquettes and contains type-I vertices only (Fig.~\ref{fig:FFs}), so that $(N_V, N_{\text{II}}) =(N,0)$ and the energy is
\begin{equation}
  E^{\text{cl}}_{\text{FF}} = V N  \,.
  \label{eq:EFF}
\end{equation}
Equating the two energies above, we get the $V=\mu$ phase boundary between the fully flippable phase and the isolated manifold.
It results in an extensively degenerate boundary carrying configurations from the side of the triangle by $N_V+N_{\text{II}} = N$. Its degeneracy may allow the quantum term to induce further phases.
The fully flippable and isolated phases appeared in the isotropic (i.e., $\mu=0$) limit of the Q6VM studied in Ref.~\cite{Shannon2004} as the doubly-degenerate N\'eel and the sub-extensively-degenerate quasi-collinear phase.

\begin{figure}
\includegraphics[width=.95\columnwidth]{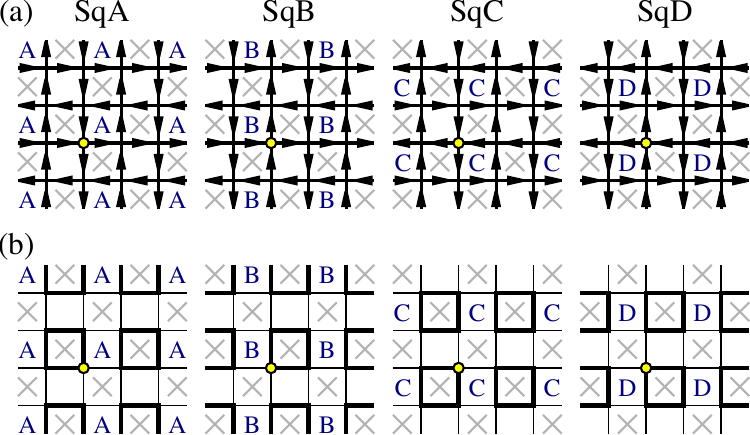}
\caption{The four square configurations made from type-II vertices in (a) the arrow and (b) the fully packed loop representation.  Compared to the fully flippable configurations in Fig.~\ref{fig:FFs},  only half of the plaquettes are flippable (grey crosses denote the non-flippable plaquettes). These four states break the translational invariance and partition the lattice into four sublattices denoted by letters A to D. We assign the letters A, B, C, and D to the plaquette in which arrows rotate counterclockwise. The site with the open circle is the same in all clusters and serves as anchor points.
\label{fig:Squares}}
\end{figure}

In addition to these known phases, we identified a third classical phase called the square phase (Fig.~\ref{fig:Squares}), where only half of the plaquettes are flippable, and all the vertices are type-II, $(N_V, N_{\text{II}})=\left({N}/{2}, N\right)$. Its energy is
\begin{equation}
  E^{\text{cl}}_{\text{Sq}} = \left(\frac{V}{2} + \mu \right)N  \,.
  \label{eq:ESq}
\end{equation}
This phase is 4-fold degenerate and breaks the translational symmetry. Comparing the $E^{\text{cl}}_{\text{Sq}}$ to $E^{\text{cl}}_{\text{FF}}$, we get the $V=2\mu$ phase boundary between the fully flippable and square phase. Similarly, $V=0$ is the classical boundary between the isolated and square phases. 

The Rys $F$ model corresponds to $V=0$. It has two phases, the "anti-ferroelectric" for $\mu>0$ and the "disordered" for $\mu<0$~\cite{Lieb1967,Sutherland1967_PhysRevLett.19.103,Baxter}. The former matches the fully flippable phase and the latter the phase boundary between the isolated and the square phases (thick red line in Fig.~\ref{fig:CPD}), with a subextensive degeneracy $W_{\text{disordered}}=4^L$ in $N=L^2$ clusters. The disordered manifold consists of all the configurations with only type-II vertices, including the square and the isolated ones. All arrows on a horizontal or vertical line point in the same direction in these configurations, the directions on different lines do not correlate. 

The classical phases, particularly square one, motivate a quadripartite division of the lattice. Selecting the position of the flippable plaquettes with a counterclockwise direction of arrows, the four square states define the A, B, C, and D kind of plaquettes, see Fig.~\ref{fig:Squares}. This partition of the plaquettes allows us to write the configurations in the fully flippable and square phases as
\begin{subequations}
    \begin{align}
       \mid \! {\rm FF1} \rangle & = \; \mid \circlearrowleft \rangle_{\rm A} \mid \circlearrowleft \rangle_{\rm D} = \; \mid \circlearrowright \rangle_{\rm B} \mid \circlearrowright \rangle_{\rm C} \; ,  \label{eq:ClassicalStatesFF1}
\\
       \mid \! {\rm FF2} \rangle & = \; \mid \circlearrowleft \rangle_{\rm B} \mid \circlearrowleft \rangle_{\rm C} = \; \mid \circlearrowright \rangle_{\rm A} \mid \circlearrowright \rangle_{\rm D} \; , \label{eq:ClassicalStatesFF2} \\
       \mid \! {\rm SqA} \rangle & = \; \mid \circlearrowleft \rangle_{\rm A} \mid \circlearrowright \rangle_{\rm D} \; , \label{eq:ClassicalStatesSqA}\\
       \mid \! {\rm SqB} \rangle & = \; \mid \circlearrowleft \rangle_{\rm B} \mid \circlearrowright \rangle_{\rm C} \; , \\
       \mid \! {\rm SqC} \rangle & = \; \mid \circlearrowright \rangle_{\rm B} \mid \circlearrowleft \rangle_{\rm C} \; , \\
       \mid \! {\rm SqD} \rangle & = \; \mid \circlearrowright \rangle_{\rm A} \mid \circlearrowleft \rangle_{\rm D} \label{eq:ClassicalStatesSqD}\;.
    \end{align}
    \label{eq:ClassicalStates}
\end{subequations}
They may serve as variational wave functions and initial states for the L\'anczos algorithm when we study the system with exact diagonalization in Sec.~\ref{sec:ED}.

So far, we discussed systems with periodic boundary conditions. Extending the results for open boundary conditions or infinite systems size requires further discussion. For example, Ref.~\cite{Zhang_2023} considers a system with domain wall boundary conditions that fix a flux sector. Quantum dynamics then splits the Hilbert space into smaller fragments within the selected flux sector (Krylov spaces). They give the number of these and show that an RK-like exact eigenstate exists in each fragmented space, among others.

\section{Symmetry groups and order parameters}
\label{sec:Symmetries}

We now turn to the $t \neq 0$ case, where the off-diagonal terms of quantum origin appear in the Hamiltonian. Previous studies determined the phase diagram as a function of $V/t$ for $\mu = 0$~\cite{Shannon2004,Banerjee_2013,Tschirsich_MPS_SciPost_2019,Ran2022}. We apply numerical and analytical approaches to extend the phase diagram with the $\mu/t$ axis. But to identify the different phases in the $(0,0)$ flux sector, we need to know the symmetries they break and the respective order parameters.
In this section, we systematically construct the order parameters from symmetry considerations using the mathematical tools of group theory. 
We conclude this section by formulating the Landau free energy and discussing the order of phase transitions.

\subsection{Symmetry groups of the different phases}
\label{sec:symm_groups}

\begin{figure}[b!]
\begin{center}
\includegraphics[width=.8\columnwidth]{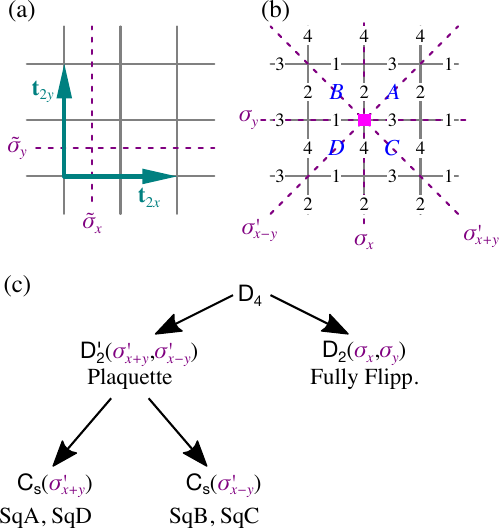}
\caption{
The elements of the symmetry group on the square lattice. 
(a) The symmetries that leave all three phases in the $(0,0)$ flux sector invariant. 
(b) The elements of the $\mathsf{G}/\mathsf{\tilde G} \cong \mathsf{D_4}$ group, their action is summarized in Table~\ref{tab:syms}. The coordinate of the A plaquette's lower left corner (magenta point) is $(0,0)$. 
(c) The symmetry groups of the different phases and the subgroup relations.
\label{fig:syms}}
\end{center}
\end{figure}

\begin{table*}[bt]
\caption{The action of the $\mathsf{D_4}$ group elements, represented by permutations, on the fully flippable, plaquette, and square states, and the six possible vertices. The bottom row shows the action of the charge conjugation operator $\mathcal{C}$. The definition of the fully flippable and the square states in the $t \rightarrow 0$ limit is provided by Figs.~\ref{fig:FFs} and \ref{fig:Squares}. In the case of the plaquette phase, AD denotes the quantum state in which the A and D plaquettes resonate, and BC, where the B and C resonate. The last five rows are the characters of the irreducible representations. We grouped the rows belonging to the same conjugacy class.}
\label{tab:syms}
\begin{ruledtabular}
\begin{tabular}{ccccccccccccrrrrr}
\multicolumn{2}{c}{$\mathsf{\mathsf{D_4}}$}  & \multicolumn{4}{c}{configurations} & \multicolumn{6}{c}{vertices}	& \multicolumn{5}{c}{irreducible repr.}	
\\
$g$  & perm.  & FF & Plaq. & \multicolumn{2}{c}{Square} & \multicolumn{2}{c}{type-I} & \multicolumn{4}{c}{type-II}&	$A_1$	&	$A_2$	&	$B_1$	&	$B_2$	&	$E$	
\\
\hline
1	&	() 			& 1	& AD	& A	& B	& \vertIa	& \vertIb	& \vertIIa	& \vertIIb	& \vertIIc	& \vertIId	&	$1$	&	$1$	&	$1$	&	$1$	&	$2$	
\\
\\
$\sigma_x$	&	(1,3) 		& 1 & BC	& B	& A	& \vertIa	& \vertIb	& \vertIIb	& \vertIIa	& \vertIId	& \vertIIc	&\multirow{2}{*}{$1$}&\multirow{2}{*}{$-1$}&\multirow{2}{*}{$1$}&\multirow{2}{*}{$-1$}&\multirow{2}{*}{$0$}	
\\
$\sigma_y$	&(2,4) 		& 1 & BC	& C	& D	& \vertIa	& \vertIb	& \vertIId	& \vertIIc	& \vertIIb	& \vertIIa	&&&&&
\\

\\
$C_2$	&(1,3)(2,4) 	& 1 & AD 	& D	& C	& \vertIa	& \vertIb	& \vertIIc	& \vertIId	& \vertIIa	& \vertIIb	&	$1$	&	$1$	&	$1$	&	$1$	&	$-2$
\\

\\
$\sigma'_{x+y}$	&(1,2)(3,4) 	& 2 & AD 	& A	& C	& \vertIb	& \vertIa	& \vertIIa	& \vertIId	& \vertIIc	& \vertIIb	&\multirow{2}{*}{$1$}&\multirow{2}{*}{$-1$}&\multirow{2}{*}{$-1$}&\multirow{2}{*}{$1$}&\multirow{2}{*}{$0$}	
\\
$\sigma'_{x-y}$	&(1,4)(2,3) 	& 2 & AD 	& D	& B	& \vertIb	& \vertIa	& \vertIIc	& \vertIIb	& \vertIIa	& \vertIId	&&&&&
\\

\\
$C_4$	&(1,2,3,4) 	& 2 & BC 	& B	& D	& \vertIb	& \vertIa	& \vertIIb	& \vertIIc	& \vertIId	& \vertIIa	&\multirow{2}{*}{$1$}&\multirow{2}{*}{$1$}&\multirow{2}{*}{$-1$}&\multirow{2}{*}{$-1$}&\multirow{2}{*}{$0$}
\\
$C_4^3$	&(1,4,3,2) 	& 2 & BC 	& C	& A	& \vertIb	& \vertIa	& \vertIId	& \vertIIa	& \vertIIb	& \vertIIc	&&&&&
\\
\hline
$\mathcal{C}$	&	& 2	& AD	& D	& C	& \vertIb	& \vertIa	& \vertIIc	& \vertIId	& \vertIIa	& \vertIIb	
\\
\end{tabular}
\end{ruledtabular}
\end{table*}%

For convenience, we work in the packed loop representation below.
All of the three phases --the fully flippable, the square, and the plaquette-- are invariant to the translations by the $\mathbf{t}_{2x}=(0,2)$ and $\mathbf{t}_{2y}=(2,0)$ lattice vectors, and the mirror symmetries $\tilde\sigma_x$ and $\tilde\sigma_y$ with vertical and horizontal axes that split the squares into half, see Fig.~\ref{fig:syms}(a). The order of the group $\mathsf{T}_2$ formed by translations $\mathbf{t}_{2x}$ and $\mathbf{t}_{2y}$ is $N/4$ in a cluster with periodic boundary conditions respecting the division into four sublattices. The two orthogonal reflections $\tilde\sigma_x$ and $\tilde\sigma_y$ generate a point group isomorphic to $\mathsf{D_2}$ with four elements. So the symmetry group $\mathsf{\tilde G} = \mathsf{D_2} \times \mathsf{T}_2$ that preserves any of these three phases has $|\mathsf{\tilde G}|  = N$ elements in the $N=L^2$ and $N=2L^2$ type clusters.

On the other hand, the Hamiltonian commutes with all the elements of the wallpaper group $\mathsf{G}$ of the square lattice, which is  $p4m$ in the  IUCr notation. 
The phases mentioned above break the symmetries of the quotient group $\mathsf{G}/\mathsf{\tilde G} \cong \mathsf{D_4}$ in one way or another. Order parameters can capture the symmetry breaking, which we will construct as irreducible representations of the $\mathsf{D_4}$. Let us mention that the quotient group $\mathsf{D_4}$ is isomorphic to the point group of the lattice, its generators are the $C_4$ rotation represented by the cyclic permutation $(1,2,3,4)$ and the reflection $(1,3)$ about the $y$ axis, following the enumeration of the edges in Fig.~\ref{fig:syms}(b). Tab.~\ref{tab:syms} lists the actions of the group elements of $\mathsf{D_4}$ on different phases and type-I and II vertices. Using this, we can identify the symmetry groups of the phases. The fully flippable states are invariant under the $\mathsf{D_2}$ group generated by the $\sigma_x$ and $\sigma_y$ reflections, with elements 
\begin{equation}
\mathsf{D_2} = \{ 1, \sigma_x, \sigma_y, C_2 \} \;.
\label{eq:D2_FF}
\end{equation}
The $\sigma'_{x+y}$ and $\sigma'_{x-y}$ reflections generate the symmetry group of the plaquette states,
\begin{equation}
\mathsf{D'_2} = \{ 1, \sigma'_{x+y}, \sigma'_{x-y}, C_2 \} \;.
\label{eq:D2_Plaq}
\end{equation}
Since there is no subgroup relation between the $\mathsf{D_2}$ and $\mathsf{D'_2}$ symmetry groups, the phase transition between the fully flippable and the plaquette phase is first order according to Landau's criterium. However, the symmetry groups
of the square phases are both subgroups of the $\mathsf{D'_2}$ group of the plaquette phase (see Fig.~\ref{fig:syms}(c)), so the transition between the plaquette and the square phases can be continuous.

\subsection{Order parameters}
\label{sec:order_pars}

To characterize the different phases, we construct order parameters below using the transformation properties of the vertices and characters, tabulated in Tab.~\ref{tab:syms}. 
It gives the transformation properties of the various ordered phases, from which we can calculate the characters and their irreducible representations. 
The fully flippable phase belongs to $A_1 \oplus B_1$ representation, the square phase to $A_1 \oplus B_2 \oplus E$, and the plaquette as $A_1 \oplus B_2$. To distinguish them, we need to construct order parameters that transform according to the $B_1$, $B_2$, and $E$ irreducible representations. Using the vertex operators, we find the following irreducible representations at a site $i$
\begin{subequations}
\label{eq:local_order_parameters}
\begin{align}
\hat o^{\text{FF}}_i &= n_i(\vertIa) - n_i(\vertIb) \;,\\
\hat o^{\text{Pl}}_i &= n_i(\vertIIa) - n_i(\vertIIb) + n_i(\vertIIc)- n_i(\vertIId) \;, \\
\mathbf{\hat o}^{\text{Sq}}_i & = \frac{1}{\sqrt{2}}
\begin{pmatrix}
  n_i(\vertIIa) + n_i(\vertIIb) - n_i(\vertIIc)- n_i(\vertIId) \\
  n_i(\vertIIa) - n_i(\vertIIb) - n_i(\vertIIc)+ n_i(\vertIId)
\end{pmatrix} \;.
\end{align}
\end{subequations}
The $\hat o^{\text{FF}}_i$ transforms as the $B_1$, the $\hat o^{\text{Pl}}_i$ as the $B_2$, and $\mathbf{\hat o}^{\text{Sq}}_i$ as the two dimensional $E$  irreducible representation of the $\mathsf{D_4}$. Here, the $n_i(\vertIa) = |\vertIa \rangle \langle \vertIa |$ is 1 if the site $i$ contains vertex $\vertIa$ and zero otherwise, and similarly for other vertices. 

Since the phases are invariant under $\mathsf{\tilde G}$, the order parameters $\hat O$ transform as the trivial irreducible representation of $\mathsf{\tilde G}$ and are given as the sum over the elements of $\mathsf{\tilde G}$ acting on the local $\hat o_i$ operators, $\hat O = \sum_{g\in \mathsf{\tilde G}} g \hat o_i$.
Performing the sum, we construct the order parameters of the various phases as
\begin{subequations}
\label{eq:order_parameters}
\begin{align}
\hat O_\text{FF} &=  \frac{1}{N}\sum_i \hat o^{\text{FF}}_i \,.
\\
\hat O_\text{Pl} &= \frac{1}{N}\sum_i (-1)^{i_x+i_y} \hat o^{\text{Pl}}_i \;,
\\
\mathbf{\hat O}_\text{Sq} & = \frac{1}{N} \sum_i
\begin{pmatrix}
(-1)^{i_y} & 0 \\
0 & (-1)^{i_x} \\
  \end{pmatrix} \cdot \mathbf{\hat o}^{\text{Sq}}_i \;.
\end{align}
\end{subequations}
The values the order parameters take in different phases are summarized in Tab.~\ref{tab:ordpar_values}.

Let us mention that
\begin{equation}
\hat o^{\text{Pl}}_i = 2 \hat o^{\text{Sq}}_{i,1} \hat o^{\text{Sq}}_{i,2} \;.
\end{equation}
Furthermore, the number operators,
\begin{subequations}
\begin{align}
n^{\text{I}}_i &= \left( \hat o^\text{FF}_i \right)^2 \;, \\
n^{\text{II}}_i &= \left( \hat o^\text{Pl}_i \right)^2 = \left( \mathbf{\hat o}^{\text{Sq}}_i \right)^2 \;,
\end{align}
\end{subequations}
are invariant under $\mathsf{D_4}$ (they transform as the $A_1$ irreducible representations).

\begin{table}[tb]
\caption{The value of the order parameters in different phases. Let us note that the $O_\text{Pl}$ is finite in both the plaquette and the square phase.}
\label{tab:ordpar_values}
\begin{ruledtabular}
\begin{tabular}{ccccccccccc}
  & \multicolumn{2}{c}{FF} & \multicolumn{2}{c}{Plaq.} & \multicolumn{4}{c}{Square}\\
$O$  & FF1 & FF2 & AD & BC & SqA & SqB & SqC & SqD \\
\hline
$O_\text{FF}$	& $-1$	& $1$	& $0$	& $0$	& $0$	& $0$	& $0$	& $0$	\\
$O_\text{Pl}$	& $0$	& $0$	& $1/2$	& $-1/2$	& $1$	& $-1$	& $-1$	& $1$	\\
$O_\text{Sq,1}$	& $0$	& $0$	& $0$	& $0$	& $1/\sqrt{2}$	& $1/\sqrt{2}$	& $-1/\sqrt{2}$	& $-1/\sqrt{2}$	\\
$O_\text{Sq,2}$	& $0$	& $0$	& $0$	& $0$	& $1/\sqrt{2}$	& $-1/\sqrt{2}$	& $1/\sqrt{2}$	& $-1/\sqrt{2}$	\\
\end{tabular}
\end{ruledtabular}
\end{table}

Eqs.~(\ref{eq:local_order_parameters}) and (\ref{eq:order_parameters})  define the order parameters using vertices, unlike the quantum dimer model, where the order parameters depend on the occupation on bonds~\cite{Sachdev_PhysRevB.40.5204_1989}. 
We may ask ourselves why cannot we follow the same construction. To this end, let us denote by $n_l$ the occupation of the bonds indexed by $l$ in Fig.~\ref{fig:syms}(b); it is one if occupied by a loop segment and -1 if not.
The operators $n_1(\dimerH)$, $n_2(\dimerV)$ ,$n_3(\dimerH)$, and $n_4(\dimerV)$ belong to the $A_1 \oplus B_1 \oplus E$ representation.   
We may write the local order parameters for the fully flippable and square phases as
\begin{subequations}
\label{eq:local_order_parameters_densities}
\begin{align}
\hat o_{\text{dimer}}^{\text{FF}} &\propto n_1(\dimerH) - n_2(\dimerV) + n_3(\dimerH) - n_4(\dimerV)\;,\\
\mathbf{\hat o}^{\text{Sq}}_{\text{dimer}} &\propto
\begin{pmatrix}
  n_2(\dimerV) - n_4(\dimerV) \\
  n_1(\dimerH) - n_3(\dimerH)
\end{pmatrix} \;.
\end{align}
\end{subequations}
However, the plaquette order parameter cannot be expressed as a linear operator in bond densities since we cannot combine $n_l$s to transform according to the $B_2$ irreducible representation. The vertex operators in Eq.~(\ref{eq:local_order_parameters}) are bilinear in bond occupation; they span a larger operator space, the $n_i(\vertIa)$ and $n_i(\vertIb)$ transforms as $A_1 \oplus B_1$ and the $n_i(\vertIIa)$,$n_i(\vertIIb)$, $n_i(\vertIIc)$, and $n_i(\vertIId)$ as $A_1 \oplus B_2 \oplus E$. The construction of the plaquette order parameter requires $B_2$. Let us mention that the staggered flippability, which is bilinear in bond occupations, is also an obvious choice for a plaquette order parameter~\cite{Syljuasen_PRL_2006, Tschirsich_MPS_SciPost_2019}. Height representation is yet another tool to construct order parameters~\cite{Banerjee_2013, Ran2022}.

To complete the analysis, the charge conjugation acts on the order parameters as
\begin{subequations}
\label{eq:C_O}
\begin{align}
\mathcal{C} \hat O_{\text{FF}} &= - \hat O_{\text{FF}} \;, \label{eq:C_O_FF}\\
\mathcal{C} \hat O_{\text{Pl}} &= \hat O_{\text{Pl}} \;,\\
\mathcal{C}
\mathbf{\hat  O}_{\text{Sq}} & = -  \mathbf{\hat O}_{\text{Sq}} \;.
\end{align}
\end{subequations}
Only the plaquette states are invariant to charge conjugation, with a symmetry group enlarged to
\begin{equation}
 \{1, \mathcal{C} \} \times \mathsf{D_2} =
 \left\{ 1, \sigma'_{x+y}, \sigma'_{x-y}, C_2, \mathcal{C}, \mathcal{C}\sigma'_{x+y},\mathcal{C}\sigma'_{x-y}, \mathcal{C}C_2 \right\},
 \label{eq:GPlaqC}
 \end{equation}
having eight elements.

The fully flippable phase breaks $\mathcal{C}$ but is invariant under the fourfold rotation combined with the charge conjugation, $\mathcal{C}C_4$. The symmetry group of this phase is then
\begin{equation}
\{1, \mathcal{C}C_4 \} \times \mathsf{D_2} =
\left\{ 1, \sigma_{x}, \sigma_{y}, C_2, \mathcal{C}C_4, \mathcal{C}\sigma'_{x+y},\mathcal{C}\sigma'_{x-y}, \mathcal{C}C_4^3 \right\}.
\label{eq:GFFC}
\end{equation}
The square phase also breaks charge conjugation, but they are invariant to the combination of the $\mathcal{C}$ and a reflection. For example, the symmetry group $\mathsf{C_s}$ of the SqA and SqD states (see Fig.~\ref{fig:syms}(c)) is extended to
\begin{equation}
 \{1, \mathcal{C}\sigma'_{x-y}\} \times \mathsf{C_s}  =  \left\{  1, \sigma'_{x+y}, \mathcal{C}\sigma'_{x-y}, \mathcal{C}C_2 \right\} .
  \label{eq:GSqC}
\end{equation}
The quotient of the groups defined in Eqs.~(\ref{eq:GPlaqC}) and  (\ref{eq:GSqC}) is isomorphic to $\mathsf{D_2}/\mathsf{C_s} \cong \mathsf{C_2}$. Even though we added the charge conjugation, a single generator remains broken at the phase transition from the plaquette phase to the square phase, preserving the possibility of a continuous transition. We will construct and analyze the Landau free energy in the next subsection to see how this happens.

\subsection{Landau free energy}
\label{subsec:Landau}

Once we identified the order parameters and their transformation properties, we can write down the free energy invariant under the $\mathsf{D}_4$ quotient group. Including up to quartic terms, its form is
\begin{align}
\mathcal{F} &=
c_{2,0,0} O_\text{FF}^2
+c_{0,2,0} O_\text{Pl}^2
+c_{0,0,2} |{\mathbf{O}}_\text{Sq}|^2
\nonumber\\&\phantom{=}
+c_{1,0,2} O_\text{FF} \left(O_{\text{Sq},2}^2- O_{\text{Sq},1}^2\right)
+c_{0,1,2} O_\text{Pl} O_{\text{Sq},1} O_{\text{Sq},2}
\nonumber\\&\phantom{=}
+c_{4,0,0} O_\text{FF}^4
+c_{0,0,4} |{\mathbf{O}}_\text{Sq}|^4
+c_{0,4,0} O_\text{Pl}^4
\nonumber\\&\phantom{=}
+c_{2,2,0} O_\text{FF}^2 O_\text{Pl}^2
+c_{2,0,2} O_\text{FF}^2 |{\mathbf{O}}_\text{Sq}|^2
+c'_{0,0,4} O_{\text{Sq},1}^2 O_{\text{Sq},2}^2
\nonumber\\&\phantom{=}
+c_{0,2,2} O_\text{Pl}^2 |{\mathbf{O}}_\text{Sq}|^2 \,.
\label{eq:Landau_full}
\end{align}
The coefficients $c$ are some functions of the couplings. 
The non-geometric charge conjugation symmetry $\mathcal{C}$ further restricts the allowed terms.
For instance, since the  $O_\text{FF}$ is odd under $\mathcal{C}$ [see Eq.~(\ref{eq:C_O_FF})], the inclusion of the $\mathcal{C}$ removes the $O_\text{FF} \left(O_{\text{Sq},2}^2- O_{\text{Sq},1}^2\right)$ term from the Landau free energy $\mathcal{F}$. We are then left with
\begin{align}
\mathcal{F} &=
c_{2,0,0} O_\text{FF}^2
+c_{0,2,0} O_\text{Pl}^2
+c_{0,0,2} O_\text{Sq}^2
\nonumber\\&\phantom{=}
+c_{0,1,2} O_\text{Pl} O_\text{Sq}^2 \sin 2\phi
+ \text{quartic terms} \,,
\label{eq:Landau_CC}
\end{align}
where we introduced the
\begin{align}
{\mathbf{O}}_\text{Sq} =
\begin{pmatrix}
  O_{\text{Sq},1} \\
  O_{\text{Sq},2}
\end{pmatrix}
=
 O_{\text{Sq}}
 \begin{pmatrix}
   \cos \phi  \\
   \sin \phi
 \end{pmatrix}
\end{align}
parametrization of the square order parameter. The appearance of the angle $\phi$ in the cubic term, together with the plaquette order parameter, is the consequence of the hierarchy of symmetry breaking presented in Fig.~\ref{fig:syms}.
For example, we can develop the SqA and SqD states from the plaquette phase resonating on the AD sublattice by breaking one of the reflections. In the Landau functional language, $\sin{2\phi}$ will be fixed to 1 throughout the phase transition between the AD plaquette and the SqA and SqD phases when the square order parameter becomes nonzero. Similarly, breaking the BC plaquette phase into SqB 
and SqC sets $\sin{2\phi} = -1$. This agrees with $\phi = \pi/4 + n\pi/2$ ($n\in \mathbb{Z}$) corresponding to the four classical square states. Moreover, it implies that $c_{0,1,2}<0$.

Because of the many undefined coefficients, it is not easy to describe the phase diagram and the order of the phase transitions.
We only mention that the Landau free energy allows both first and second-order transition for the phase boundary between the square and the plaquette phase, depending on the sign of the quartic term. In Appendix \ref{sec:variational_WF}, we present a simple variational wave function to describe the plaquette-square transition. It displays both a first and a continuous boundary separated by a tricritical point.

Let us note that the charge conjugation symmetry is particular for the 6-vertex model, as half of the bonds are occupied in the fully packed loop representation.
Hence, the charge conjugation symmetry is absent in the quantum dimer model.

\section{The manifold of the isolated states}
\label{sec:isolated}

\begin{figure}[bt]
\includegraphics[width=.95\columnwidth]{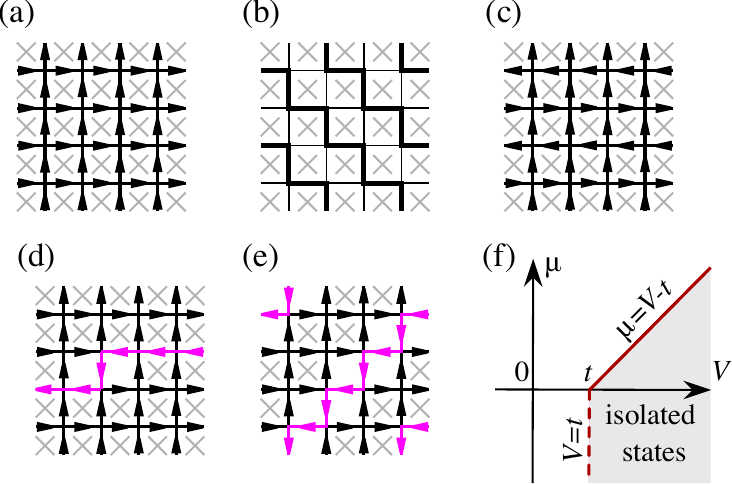}
\caption{\label{fig:iso_states}
The isolated configuration in the $\mathbf{m} = (L, L)$ flux sector in (a) the arrow and (b) the fully packed loop representation. (c) An isolated configuration from the  $\mathbf{m} = (0, L)$ flux sector. Flippable plaquettes are absent, and all the vertices are type-II ($N_V=0$ and $N_{\text{II}}=N$). The order of the left- and right-pointing horizontal lines is arbitrary, so the number of states in this flux sector is $\binom{4}{2}=6$.
(d) A mobile leapfrog excitation (`frogton') with two flippable plaquettes ($N_V=2$, $N_{\text{I}}=2$) in a cluster with suitable cluster geometry.
(e) A state in the $\mathbf{m} = (L-2,L-2)$ flux sectors. It results from reversing the arrows in the isolated state shown in (a) along the closed magenta loop crossing the boundary. 
(f) The minimal extent of the isolated state manifold. The $\mu = V-t$ boundary (solid line) denotes the instability line against frogtons. The isolated states, however, can extend beyond the dashed boundary at $V=t$.}
\end{figure}

Configurations without flippable plaquettes form the isolated manifold. They are disconnected from other configurations by local flips and consist only of type-II vertices if periodic boundary conditions are imposed.
 The quantum term $\mathcal{H}_t$ [Eq.~(\ref{eq:Ht})] annihilates any of them, so their energy remains the classical one, $E_{\text{Iso}} = \mu N$. 

Ref.~\cite{Shannon2004} provides a recipe to construct all the configurations in the isolated manifold. 
If all the arrows along the horizontal or vertical lines in a configuration point in the same direction, with the proviso that either the horizontal or the vertical lines must be oriented alike, none of the plaquettes is flippable. See Fig.~\ref{fig:iso_states}(a) and (c) for examples. It implies that at least one of the flux indices of these isolated configurations has to be extremal. 

This recipe helps to determine the degeneracy of this manifold. Let us consider the $N=L^2$ cluster as an example. We then have two choices of fixed orientation, both in the case of horizontal and vertical lines, resulting in a factor of four (Fig.~\ref{fig:iso_states}(a) illustrates one of these four states). In the non-fixed direction, each of the $L$ lines can point in two directions, giving $2^L$ possibilities altogether. 
Supposing that $l$ lines point in one and $L-l$ in the other direction defines configurations in the $(\pm L, L-2l)$ or $(\pm L, L-2l)$ flux sectors, with degeneracies $\left(L \atop l\right)$.
 Considering the double counting of $(\pm L, \pm L)$ flux sectors, we end up with $4 \times 2^L - 4$ isolated states in the manifold. The degeneracy exponentially grows with the linear size of the system, so it is subextensive.
We can apply similar considerations to clusters with other geometries. 

Knowing their energies allows us to determine an exact region in the phase diagram where they are the ground states. For this purpose, we will use Gerschgorin's theorem, which claims that $| H_{ii} - \varepsilon_{i} | \leq \sum_{j \neq i} | H_{ij} | $ for a finite cluster Hamiltonian $\mathcal{H}$ with eigenvalues $\{ \varepsilon_i \}$.
We use a basis where both $\hat N_V$ and $\hat N_{\text{II}}$ operators are diagonal, and the quantum term $\mathcal{H}_t$ is strictly off-diagonal. Then for the $i$th configuration having $N_V$ flippable plaquettes and $N_{\text{II}}$ type-II vertices the diagonal term is $H_{ii} = V N_V + \mu N_{\text{II}}$. The sum over the off-diagonal terms  $\sum_{j \neq i} | H_{ij} | = N_V t$, as the configuration connects to exactly $N_V$ other ones, each with $-t$ amplitude (note that we chose $t>0$). So we can write
\begin{equation}
| V N_V + \mu N_{\text{II}} - \varepsilon_{i} | \leq   N_V t \,.
\end{equation}
Resolving the absolute value, the $\varepsilon_{i}$s become bounded as
\begin{equation}
 (V-t) N_V + \mu N_{\text{II}}  \leq \varepsilon_{i}  \leq (V +t)  N_V + \mu N_{\text{II}} \,.
\end{equation}
Let us denote by $\Delta_i = \varepsilon_{i} - \mu N$ the energy gap between the eigenvalue $\varepsilon_{i}$ and the energy of the isolated manifold. Then
\begin{equation}
 (V-t) N_V + \mu (N_{\text{II}}-N)  \leq \Delta_{i} \,.
\end{equation}
Isolated states are ground states while the gap $\Delta_{i} \geq 0$, which is satisfied when $(V-t) N_V - \mu (N - N_{\text{II}})  \geq 0 $. We need to find the region in the parameter space of $V$, $\mu$, and $t$, where this inequality holds, provided that Eqs.~(\ref{eq:GeomConstrains})  constrain the possible $N_V$ and $N_{\text{II}}$ values into the triangle shown in Fig.~\ref{fig:nVnII_map}. It is a simple linear optimization problem analogous to finding the classical phase diagram with $V$ replaced by $V-t$. The extrema occur at the corners of the triangle, where $(N_V, N_{\text{II}}) = (0, N)$ (isolated phase), $(N/2, N)$ (square phase), or $(N,0)$ (fully flippable phase). Eventually, we conclude that the isolated configurations form the ground state manifold when
\begin{equation}
  0 \leq \mu \leq V-t \quad \text{or} \quad \mu \leq  0 \leq V-t \,.
  \label{eq:isoBoundary}
\end{equation}
Fig.~\ref{fig:iso_states}(f) shows this region. The approach above does not tell us whether the isolated states remain the ground states outside this region. Below we will consider excitations that become soft at the positive $\mu = V-t $ boundary of the region above, promoting it to a phase boundary.
\begin{figure}[b!]
\begin{center}
\includegraphics[width=.9\columnwidth]{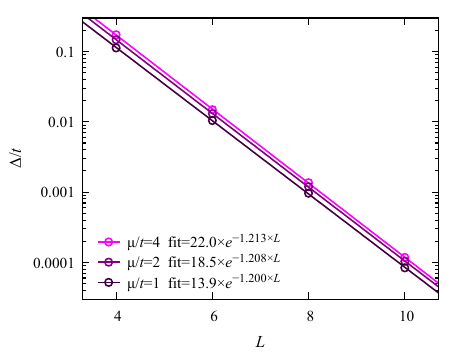}
\caption{The finite size scaling of the gap $\Delta = E_{\text{GS}}^{(L-2,L-2)} - E_{\text{Iso}}$ between the ground state energy in the $\mathbf{m}=(L-2,L-2)$ flux sector and the $E_{\text{Iso}}=N \mu$ of the isolated state on the $V=\mu+t$ line in the phase diagram. 
 We show three selected cases ($\mu/t=1$, $2$, and $4$) in $N=L^2$ class of clusters, calculated by the L\'anczos method.
The size of the Hilbert space is 90 for $L=4$ [see Fig.~\ref{fig:flux_16}], 2\,772 for $L=6$ [see Fig.~\ref{fig:topsec_32_36}(b)], 51\,480 for $L=8$, and 923\,780 for $L=10$ in the $(L-2,L-2)$ flux sectors.
 The gap exponentially vanishes with the linear size of the cluster $L$ (note the logarithmic vertical axis), and the solid lines show the fit to the $a e^{-b L}$ function. The $b$ values depend weakly on the value of $\mu$.  
\label{fig:gap_RKline}}
\end{center}
\end{figure}

Following Ref.~\cite{Shannon2004}, we can calculate the energy of a leapfrog excitation (we call it `frogton'), depicted in Fig.~\ref{fig:iso_states}(d). It exists in a cluster of a suitable skew shape, defined by the $g_1 = (L,2)$ and $g_2 = (0, L)$ lattice vectors, for example. The boundary condition introduces a correlated kink-antikink pair on the line along which the arrows are reversed [the magenta line in Fig.~\ref{fig:iso_states}(d)]. A frogton comprises two flippable plaquettes and two type-I vertices, so the diagonal energy is $2(V-\mu)$. Flipping one of the two plaquettes, the excitation hops with an amplitude $-t$ and acquires a bandwidth of $4t$ centered at the diagonal energy. The minimal energy of this variational state gives an upper bound for the gap, $\Delta = 2(V-\mu-t)$. Combined with Eq.~(\ref{eq:isoBoundary}), the gap closing along the line 
\begin{equation}
V = \mu + t
\label{eq:iso_exact_boundary}
\end{equation}
for $\mu\geq 0$ in the phase diagram gives the exact boundary of the isolated phase. 
  
In a square-shaped cluster with $N=L^2$ geometry, the flux sector that closes the gap originates from the "staircase-like" excitation shown in  Fig.~\ref{fig:iso_states}(e) in the $\mathbf{m} = (L-2, L-2)$ flux sector. 
Unlike frogtons, where the kink-antikink pair is confined within a lattice spacing, in this cluster, the number of the kinks and antikinks is not conserved, making the calculations more complicated. 
Exact diagonalizations up to $L=10$ in $N=L^2$ clusters reveal that the finite size gap between the ground state energy in the $ (L-2, L-2)$ flux sector and the energy of the isolated manifold exponentially decreases with the system size along the $V=t+\mu$ line, as demonstrated in Fig.~\ref{fig:gap_RKline}. Therefore we may conclude that the gap closes along the whole $V=t+\mu$ line in the thermodynamic limit. We will further scrutinize this question in Sec.~\ref{sec:RK}. 

\section{Phase diagram from exact diagonalization }
\label{sec:ED}

\begin{figure}[b!]
\begin{center}
\includegraphics[width=.9\columnwidth]{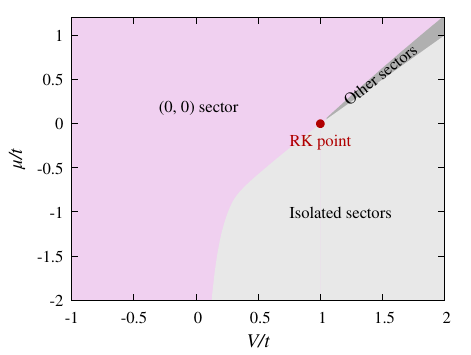}
\caption{The ground state flux sectors in the $32$ site cluster. Besides the $\mathbf{m}=(0,0)$ (magenta) and isolated sectors (white), other sectors emerge in a tiny region fanning out from the RK point (the gray area). We will discuss them in  Sec.~\ref{sec:rainbow}.
\label{fig:GS_secors}}
\end{center}
\end{figure}

\begin{figure}[t!]
\includegraphics[width = 0.95\columnwidth]{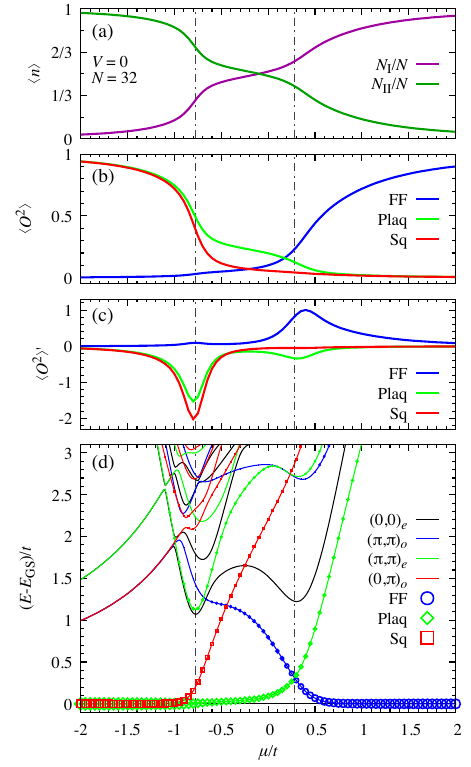}
\caption{Exact diagonalization results for the 32-site cluster at $V=0$ as a function of $\mu/t$. (a) The density of the type-I ($n_{\text{I}}=N_{\text{I}}/N$) and II ($n_{\text{II}}=N_{\text{II}}/N$) vertices in the ground state. (b) The expectation values of the squared order parameters and (c) their numerical derivatives with respect to $\mu/t$. We identify inflection points 
of the $\langle \tilde O^2 \rangle$ (i.e., the extrema of the $\langle \tilde O^2 \rangle'$) with phase boundaries. (d) The low-energy excitations are defined by their momentum and charge conjugation parity. The size of the symbols indicates the overlap between the initial states given by Eqs.~(\ref{eq:SpecStates}) and the energy eigenstates. The ground state is fully symmetric. The level crossing of the lowest lying $(\pi,\pi)_e$ and $(\pi,\pi)_o$ symmetry levels at $\mu/t\approx 0.29$ serves as an alternative indicator of the phase boundary between the plaquette ($ -0.78 \lesssim \mu/t\lesssim 0.29$) and the fully flippable phases ($\mu/t\gtrsim 0.29$). Avoided level crossings around $\mu/t \approx -0.78$ in the first excitations of the $(0,0)_e$ and $(\pi,\pi)_e$ symmetry sectors indicate the boundary between the square ($\mu/t\lesssim -0.78$) and the plaquette phase.
\label{fig:GUS32}}
\end{figure}

\begin{figure}[t!]
\includegraphics[width = 0.95\columnwidth]{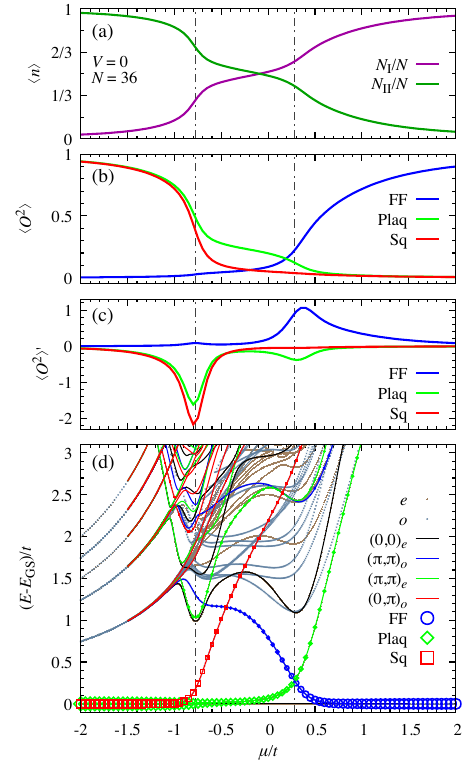}
\caption{The same as Fig.~\ref{fig:GUS32} for $N=36$. The positions of the phase transitions are unchanged compared to the 32-site cluster. We also show all the states in the selected energy window, indicating their $\mathcal{C}$ parity. 
\label{fig:GUS36}}
\end{figure}

The number of the ice rule configurations grows exponentially with the system size $N$, restricting the size of the clusters that can be studied numerically. Using the conservation of the fluxes $(m_x,m_y)$, we diagonalized 16, 32, and 36 site clusters having the full $\mathsf{D_4}$ point group symmetry of the square lattice. We used the standard L\'anczos algorithm to calculate the energy of the ground state and low-lying excitations.

To begin, we scanned the $V/t$ and $\mu/t$ parameter space to reveal which flux sectors give the ground state in the $N=16$ and $32$ site clusters. 
Fig.~\ref{fig:GS_secors} illustrates our findings.
For large values of $|V/t|$ and $|\mu/t|$,  we recovered the $(0,0)$ flux sector and the isolated manifold, as anticipated based on the classical phase diagram. 
The lowest energies become equal in all the flux sectors at the $V=t$ Rokhsar-Kivelson (RK) point. 
Fig.~\ref{fig:GS_secors} also reveals that the RK point is multicritical, {\it i.e.}, several phases merge at the RK point. 
In addition to the predicted flux sectors, a narrow sliver (gray region in Fig.~\ref{fig:GS_secors}) emanates from the RK point with monotonically varying flux sectors from the $\mathbf{m}=(0,0)$ and the isolated manifold for positive $\mu$ values. 
Computations on $36$ site clusters also confirmed the existence of this phase. In this section, we discuss the phases in the (0,0) and isolated sectors, and we will present results about the grey region in Sec.~\ref{sec:rainbow}.

Identification of the isolated phase is numerically straightforward. Since we know the exact energy of the isolated states, $E_\text{Iso} = \mu N$, it is enough to calculate the lowest energy levels of the non-isolated flux sectors with ED and compare them to $E_\text{Iso}$. Distinguishing the different phases in the (0,0) flux sector is more challenging. We applied two methods, (i) one based on the order parameters and (ii) one based on energy level spectroscopy~\cite{Shannon2004,Banerjee_2013}.

Re (i), we computed the expectation values of the squares of the order parameters given in Eqs.~(\ref{eq:order_parameters}), $\langle \text{GS} |\hat{O}_{\rm FF}^2 |\text{GS}\rangle$, $\langle \text{GS} | \hat{O}_{\rm Plaq} ^2 |\text{GS}\rangle$, and $\langle \text{GS}| \hat{\textbf{O}}_{\rm Sq}\cdot\hat{\textbf{O}}_{\rm Sq} |\text{GS} \rangle$, by calculating the ground state wave function $|\text{GS}\rangle$ numerically. We show the result along the $V=0$ line in Figs.~\ref{fig:GUS32}(b) and \ref{fig:GUS36}(b) for the 32 and 36 site clusters. 
Since the squares of the order parameters transform as the $A_1$ irreducible representation in Tab.~\ref{tab:syms} and contribute to the energy [c.f. the Landau functional in Eq.~(\ref{eq:Landau_CC})], we estimate the phase boundaries by inflection points of their expectation values. For this purpose, we calculate the extrema of the $\partial \langle \hat{O}^2 \rangle/\partial (\mu / t)$ in Figs.~\ref{fig:GUS32}(c) and \ref{fig:GUS36}(c).
According to this criterium, the square phase is realized for $\mu/t \lesssim -0.78$, the plaquette phase between $-0.78 \lesssim \mu/t \lesssim 0.3-0.4$, and the fully flippable phase for $0.3-0.4 \lesssim \mu/t$  (the reason behind the uncertainty for the upper boundary is that the inflection points of the 
$\langle \text{GS} |\hat{O}_{\rm FF}^2 |\text{GS}\rangle$ and $\langle \text{GS} | \hat{O}_{\rm Plaq} ^2 |\text{GS}\rangle$ are not at the same $\mu/t$ values).
 We note that the plaquette order parameter is also finite in the square phase, as expected from the discussion of the hierarchy of the symmetry breaking in Sec.~\ref{sec:order_pars}.

Re (ii), energy level spectroscopy provides a less direct but more powerful tool to estimate the phase boundaries in a finite-size cluster~\cite{Sindzingre_PhysRevB.66.174424}. 
Since the Hamiltonian preserves all the symmetries of the model, its wave functions transform as the irreducible representations of the underlying group. 
We can identify a set of low-lying states in the spectra, belonging to specific irreducible representations of the $\mathsf{D_4}$ quotient group. 
They collapse into a degenerate ground state manifold in the thermodynamic limit.
The symmetry-breaking states are linear combinations of the wave functions in this manifold.
 Characterization of these low-lying states gives the basis for level spectroscopy, a tool for detecting the various phases.

As an example, let us consider the fully flippable phase. 
The classical $|\text{FF}1\rangle$ and $|\text{FF}2\rangle$ configurations manifestly break the $\sigma'_{x+y}$ reflection symmetry, but we may linearly combine them into the
\begin{subequations}
\label{eq:SpecStatesFF}
\begin{align}
\mid \! \text{FF}(0,0)_e \rangle & = \; \mid \! \text{FF1} \rangle \; + \mid \! \text{FF2} \rangle \; , \\
\mid \! \text{FF}(\pi,\pi)_o \rangle & = \; \mid \! \text{FF1} \rangle \; - \mid \! \text{FF2} \rangle \;. \label{eq:SpecStatesFF_ppo}
\end{align}
\end{subequations}
For a finite cluster, the energy of the $(0,0)_e$ level is different from the $(\pi,\pi)_o$ level, but we expect the gap between them to vanish in the thermodynamic limit when the fully flippable phase is realized. 

We can associate the linear combinations above with the irreducible representations listed in Tab.~\ref{tab:syms}. More generally, the $(0,0)_e$ belongs to the trivial $A_{1,e}$ irreducible representation, the $(\pi,\pi)_e$ transforms as $B_{2,e}$, the $(\pi,\pi)_o$ transforms as the $B_{1,o}$, and the $(\pi,0)_o$ and $(0,\pi)_o$ span the two-dimensional $E_o$. Since the charge conjugation $\mathcal{C}$ commutes with all the point group elements in $\mathsf{D_4}$, the irreducible representation of the $\{1,\mathcal{C}\}\times \mathsf{D_4}$ group are simply the irreducible representation of the $\mathsf{D_4}$ appended with the even ($e$) or odd ($o$) parity with respect to $\mathcal{C}$. We note that one shall be careful with the interpretation of the momentum labels, as they can be different for the arrow and fully packed loop representation (for example, the fully flippable state is translationally invariant in the fully packed loop representation but not in the arrow representation). Above, we used the arrow representation.

We prepare states with appropriate momentum and parity and use them as input to the L\'anczos code since the iterations in the method preserve their symmetry.  
 Figs.~\ref{fig:GUS32}(d) and \ref{fig:GUS36}(d) show the momentum and parity resolved spectra for the most important $(0,0)_e$, $(\pi,\pi)_e$, $(\pi,\pi)_o$, and $(\pi,0)_o$ symmetry sectors [$(\pi,0)_o$ and $(0,\pi)_o$ are degenerate]. 
The ground state is always in the $(0,0)_e$ sector. For $\mu/t \gtrsim 0.29$, the first excited state is in the $(\pi,\pi)_o$ sector, just like in the decomposition (\ref{eq:SpecStatesFF_ppo}), and it becomes degenerate with the ground state as $\mu/t$ increases and approaches the classical limit. For $\mu/t \lesssim 0.29$, the first exited state is in the $(\pi,\pi)_e$ sector, and a level crossing occurs at $\mu/t \approx 0.29$. Since the phases break discrete symmetries, their spectrum is gapped. 
The low-lying excitations belonging to different irreducible representations on the two sides of the level crossing lead to different symmetry breaking and, thus, phases in the thermodynamic limit. 
We identify the phase for $\mu/t \gtrsim 0.29$ as fully flippable. 
Ideally, a finite-size scaling should be performed to accurately determine the phase boundary, as in Ref.~\cite{Shannon2004} for the case $\mu=0$.
However, the positions of the level crossings in $1/N$ are not monotonic for general values of the parameters. 
Nevertheless, our value agrees well with the result of quantum Monte Carlo simulation \cite{Syljuasen_PRL_2006}, $t/\mu \approx 3.47$ (i.e., $\mu/t \approx 0.288$).

\begin{figure}[b!]
\begin{center}
\includegraphics[width=.9\columnwidth]{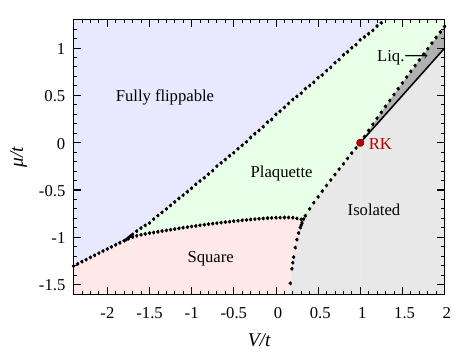}
\caption{The ground state phase diagram based on ED for the $32$-site cluster. The phase boundaries agree with the $N=36$ result up to two decimal places. We present the diagram for the smaller systems because of the better resolution. The points denote the positions of the (avoided) level crossings and inflection points of the squared order parameters. 
The boundary between the phase with finite flux sectors and the isolated states (solid line) is the exact one of  Eq.~(\ref{eq:iso_exact_boundary}). The red point denotes the Rokshar-Kivelson (RK) point.  
\label{fig:ED_phase_diag}}
\end{center}
\end{figure}

Next, consider the limit for negative $\mu$, where the square phase appears. The combinations of the four classical square configurations give the following momentum and parity eigenstates:
\begin{subequations}
\begin{align}
\mid \! \text{Sq}(0,0)_e \rangle & = \; \mid \! \text{SqA} \rangle \; + \mid \! \text{SqB} \rangle \; + \mid \! \text{SqC} \rangle \; + \mid \! \text{SqD} \rangle \; , \\
\mid \! \text{Sq}(\pi,\pi)_e \rangle & = \; \mid \! \text{SqA} \rangle \; - \mid \! \text{SqB} \rangle \; - \mid \! \text{SqC} \rangle \; + \mid \! \text{SqD} \rangle \; , \\
\mid \! \text{Sq}(\pi,0)_o \rangle & = \; \mid \! \text{SqA} \rangle \; - \mid \! \text{SqB} \rangle \; + \mid \! \text{SqC} \rangle \; - \mid \! \text{SqD} \rangle \;, \\
\mid \! \text{Sq}(0,\pi)_o \rangle & = \; \mid \! \text{SqA} \rangle \; + \mid \! \text{SqB} \rangle \; - \mid \! \text{SqC} \rangle \; - \mid \! \text{SqD} \rangle \;.
\end{align}
\label{eq:SpecStates}
\end{subequations}
In Figs.~\ref{fig:GUS32}(d) and \ref{fig:GUS36}(d), four states with precisely these momenta and parities are quasi-degenerate for $\mu/t \lesssim -0.78$, supporting the realization of the square phase in this region. As $\mu$ increases from $\mu/t \lesssim -0.78$, the energy levels of the $(0,\pi)_o$ and   $(\pi,0)_o$ split off, and only the $(\pi,\pi)_e$ remains quasi-degenerate with the ground state. They constitute the two plaquette states, with resonant A and D or B and C plaquettes, represented with the
\begin{subequations}
\begin{align}
  | \text{PlAD} \rangle &= 
   \prod_{\text{A,D plaquettes}}  
   \frac{| \! \circlearrowleft \rangle_{\rm} + | \! \circlearrowright \rangle_{\rm}}{\sqrt{2}} \;,
    \label{eq:Plaq_WFAD}\\
  | \text{PlBC} \rangle &= 
   \prod_{\text{B,C plaquettes}}  
   \frac{| \! \circlearrowleft \rangle_{\rm} + | \! \circlearrowright \rangle_{\rm}}{\sqrt{2}} \;.
 \label{eq:Plaq_WFBC}
\end{align}
\end{subequations}
approximate wave functions.
The $|\text{PlAD} \rangle  - |\text{PlBC} \rangle $ is in the $(\pi,\pi)_e$ symmetry sector.
Instead of a level crossing, an avoided level crossing characterizes the square-plaquette phase transition.

To further elucidate the nature of the excitation spectrum, we calculated the dynamical correlation functions of the order parameters,
\begin{equation}
   S^{\alpha}(\omega) = \sum_X \left|\langle X | \hat O_\alpha | \text{GS} \rangle \right|^2 \delta(\omega - E_X + E_{\text{GS}}) \;,
   \label{eq:S_alpha_omega}
\end{equation}
using the L\'anczos method, where $X$ are excited states and $\alpha=\text{FF}, \text{Sq}, \text{Pl}$ denotes the order parameter. We first calculated the ground state $| \text{GS} \rangle$ for a given value of parameters $\mu$ and $V$, applied the operator $\hat O_\alpha$ to the $| \text{GS} \rangle$, and then used $\hat O_\alpha | \text{GS} \rangle$ as the initial state for the second run of the L\'anczos procedure. The algorithm then computes the matrix elements in the definition of $S^{\alpha}(\omega)$. Not surprisingly, the largest matrix elements are for the lowest-lying excitations of the momenta and parities corresponding to the symmetry of the phase. 
In Figs.~\ref{fig:GUS32}(d) and \ref{fig:GUS36}(d), the size of the open symbols is proportional to the values of the matrix elements.

Using the above criteria, we established the phase diagram in the parameter space of $V/t$ and $\mu/t$, Fig.~\ref{fig:ED_phase_diag}. 
We determined the first-order boundaries between the fully flippable and the plaquette and between the fully flippable and the square phases by following the positions of level crossings and the boundary between the square and the plaquette phase following the positions of the inflection points.
We also checked that the parameter values of the avoided level crossings coincide with those of the inflection points. We obtained the boundary to the isolated state manifold by comparing their energy to the ground state energies in the $(0,0)$ flux sectors.  

The main consequence of the quantum fluctuations is the appearance of the plaquette phase with resonating alternating plaquettes that fills up the central region of the phase diagram. The plaquette phase extends along the $V\approx \mu$ line to larger positive values of $V$, following the unknown phase with finite flux sector ground states that separate it from the isolated states. Otherwise, for large $|V|$ and $|\mu|$ values, the phase diagram is consistent with the classical one, shown in Fig.~\ref{fig:CPD}. In the next section, we will confirm the validity of some of these phase boundaries using perturbation theory.

\section{Phase boundaries from the perturbation theory}
\label{sec:perturb}

\begin{figure}[bt!]
        \centering
        \includegraphics[width = 0.9 \columnwidth]{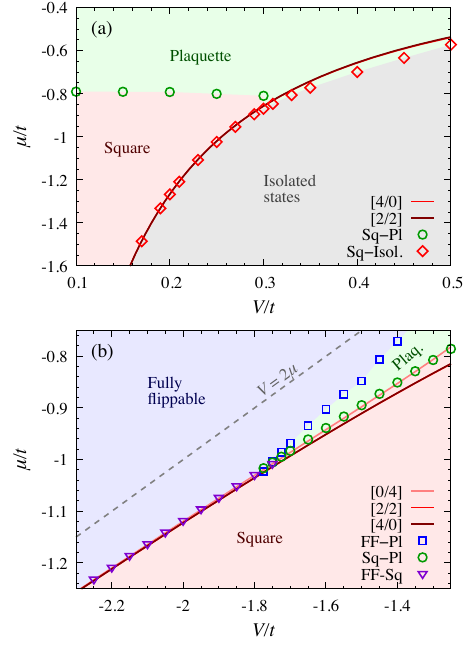}\\
    \caption{\label{fig:PertPhaseDiags}
Comparison of the phase boundaries calculated by ED for a 32-site cluster and by the fourth-order perturbation expansion in $t$. We show the Pad\'e approximants of the perturbation series to estimate its convergence. 
(a) ED data and Pad\'e approximants of the boundary between the square and the isolated phases near the triple point. The coordinates for the triple point are cluster dependent, for $N=32$  $V/t = 0.32$ and $\mu / t = -0.83$, while for $N = 36$ are $V /t = 0.29$ and $\mu /t = -0.87$. 
(b) The perturbational curve gives the boundary between the fully flippable and the square phases. It meets the lines (FF-Pl and Sq-Pl) from the ED at the triple point $V/t\approx-1.75$ and $\mu/t \approx -1.01$. The different Pad\'e approximants do not deviate significantly in the relevant $V/t \lesssim -1.75$ range. The grey dashed line shows the $V=2\mu$ classical phase boundary.
}
\end{figure}

We use the size-consistent Rayleigh-Schrödinger perturbation theory below to estimate the ground state energies in the fully flippable and square phases. We calculate the second- and fourth-order corrections in $\mathcal{H}_t$ to the ground state energy of the configurations drawn in Figs.~\ref{fig:FFs} and \ref{fig:Squares} in the $t\to0$ limit. We get an estimate for the first-order phase boundary between these phases by comparing their energies. Furthermore, comparing the energy of the square phase to $E_\text{Iso}=\mu N$ will provide the corresponding phase boundary.

The perturbation series calculation is straightforward for the square and fully flippable phases in any representation in which the $\mathcal{H}^{\rm cl}$ is diagonal. We give details in Appendix \ref{sec:perturb_appendix}.
The ground state energies up to the fourth order in the flipping amplitude $t$ are
\begin{subequations}
\begin{align}
   \frac{E_{\rm FF}}{N} & = V
   + \dfrac{t^2}{4(V - \mu)}
   \nonumber\\ & \phantom{=}
   + \dfrac{t^4}{16(V-\mu)^2} \left[ \dfrac{8}{7V - 8\mu} - \dfrac{23}{12(V-\mu)} \right]
   + \cdots \;, \\
   \frac{E_{\rm Sq}}{N} & = \dfrac{V}{2} + \mu
   + \dfrac{t^2}{8\mu}
   + \dfrac{t^4}{16\mu^2} \left[ \dfrac{4}{8\mu - V} - \dfrac{11}{24\mu} \right] + \cdots \;.
\end{align}
\end{subequations}
Solving the $E_{\rm FF} = E_{\rm Sq}$, we get the following Pad\'e approximants
for the phase boundary between the fully flippable and the square phases,
\begin{equation}
V = 2 \mu  \times
\begin{cases}
1 -\frac{t^2}{8 \mu^2} -\frac{5 t^4}{384 \mu^4}
& \text{Pad\'e} [4/0],\\
\left(1-\frac{11 t^2}{48 \mu^2}\right)
\left(1-\frac{5 t^2}{48 \mu^2}\right)^{-1}
& \text{Pad\'e} [2/2],\\
\left(1 + \frac{t^2}{8 \mu ^2} + \frac{11 t^4}{384 \mu ^4}\right)^{-1} & \text{Pad\'e} [0/4].\\
\end{cases}
\end{equation}
All three Padé approximants equally satisfy the energy equation up to the fourth order. The orders $[m/n]$ of the Padé approximants above denote the power of $t$ in the numerator ($m$) and the denominator ($n$).

Similarly, the $E_{\rm Sq} = E_\text{Iso}$ equation provides the phase boundary between the square and the isolated phases. The Pad\'e approximants are
\begin{equation}
 V = -\dfrac{t^2}{4 \mu}  \times
\begin{cases}
\left( 1 -\frac{t^2}{48 \mu^2}  \right)  & \text{Pad\'e} [4/0],\\
\left( 1 +\frac{t^2}{48 \mu^2}  \right)^{-1} & \text{Pad\'e} [2/2] . \\
\end{cases}
\end{equation}
Fig.~\ref{fig:PertPhaseDiags} shows these approximants together with the numerical results of the ED calculation for both phase boundaries. 
The comparison of different orders of Pad\'e approximants allows us to estimate the convergence of the perturbation series: the different lines are essentially superimposed on each other in the relevant domains, indicating a rapid convergence of the series. The perturbation expansion also agrees well -- typically within two decimal places -- with the phase bounds extracted from ED calculations on finite clusters.

\section{The Rokhsar-Kivelson point and the liquid phase}
\label{sec:rainbow}

The exact diagonalization of the 32-site cluster shows that the flux sector of the ground state gradually increases from the $\mathbf{m} = (0,0)$ in the plaquette phase as we approach the manifold of isolated states (the dark grey area in Fig.~\ref{fig:GS_secors}).
To gain a deeper insight into the properties of this phase, wedged between the plaquette phase and the isolated states and emanating from the quantum critical RK point, we perform a Monte Carlo evaluation of the RK wave function \cite{Henley_2004,Hermele2004}.  This enables the extension of cluster sizes to up to 576 sites close to the RK point.

\subsection{First order perturbation around the Rokshar-Kivelson point}
\label{subsec:MonteCarlo}

The RK point is a particular point in the phase diagram since the exact ground state wave function $| \text{RK}({\mathbf{m}}) \rangle$ is known: it is the equal amplitude superposition of the configurations in an $N_V$ diagonal basis within the flux sector $\mathbf{m}$ \cite{RokhsarKivelson1988}. The ground state energy is the same in all the flux sectors. But the expectation values of the flippable plaquettes $\hat N_V$ and $\hat N_{\text{II}}$ operators depend on $\mathbf{m}$.
We use the Hellman-Feynman theorem at the RK point to estimate the splitting of the ground states. In the first order, we approximate the lowest energy of the perturbed RK Hamiltonian in a given flux sector with the formula
\begin{equation}
    E_{\mathbf{m}} =
    (V - t) \langle N_{V} \rangle_{\mathbf{m}} +
    \mu \langle N_{\rm II} \rangle_{\mathbf{m}}
    \,.
    \label{eq:E_RK_1storder}
\end{equation}
The
\begin{subequations}
\begin{align}
  \langle N_{V} \rangle_{\mathbf{m}} =
  \langle \text{RK}({\mathbf{m}})| \hat N_{V}| \text{RK}({\mathbf{m}}) \rangle  \,,
  \label{eq:RK_NV_exp}
\\
  \langle N_{\rm II} \rangle_{\mathbf{m}} =
  \langle \text{RK}({\mathbf{m}})| \hat N_{\rm II}| \text{RK}({\mathbf{m}}) \rangle \,
\label{eq:RK_NII_exp}
\end{align}
\end{subequations}
denote the expectation values of the number of flippable plaquettes and type-II vertices. Comparing these energies, we can figure out the flux sector of the ground state (a similar argument appeared in Ref.~\cite{Moessner_PRL_2001} for the quantum dimer model on the triangular lattice and in Ref.~\cite{PhysRevLett.115.217202} for a quantum dimer model on the honeycomb lattice).

\subsection{The Monte Carlo method}

The Monte Carlo method uses the RK wave function to evaluate the expectation values by random sampling~\cite{Hermele2004}.
We started the simulation from a configuration formed by arrows directed along horizontal and vertical lines since this allowed the selection of the flux sector and generated new configurations by randomly flipping plaquettes. We discarded the first 5 million configurations to reach thermalization. Following thermalization, we measured $N_{\rm II}$ and $N_V$ after every $N$ step and updated their averages, where $N$ is the system size. The number of elementary steps in a Monte Carlo run was typically between $5\times 10^7$ and $10^9$ flips, depending on the system size and the statistical error. After we exported the averages and continued measuring another four times.
Repeating the procedure above five times, we collected  $N_{\text{MC}}=25$ average value pairs for each flux sector. Denoting by $m_i$  ($i=1,\dots, N_{\text{MC}}$) the averages (means) from the Monte Carlo runs, we estimate the statistical error by the standard error of the mean, given by the formula
\begin{equation}
  \sigma = \sqrt{
   \frac{\sum_{i=1}^{N_{\text{MC}}} (m_i - \overline{m} )^2}
   {N_{\text{MC}} (N_{\text{MC}} - 1)}
   } \;.
\end{equation}
Here the
\begin{equation}
  \overline{m} = \frac{1}{N_{\text{MC}}}
  \sum_{i=1}^{N_{\text{MC}}} m_i
\end{equation}
is the mean value of the $m_i$ averages.

\subsection{Dependence of expectation values on flux sectors}

\begin{figure}[tb!]
        \centering
\includegraphics[width = 0.9 \columnwidth]{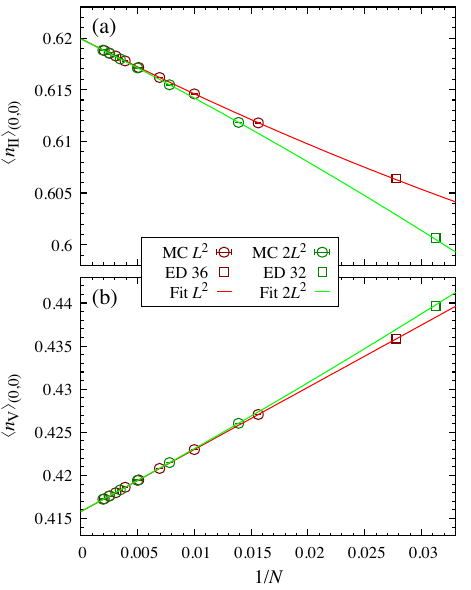}
    \caption{Finite-size scaling of the densities of (a) type-II vertices $n_{\text{II}}$ and (b) flippable plaquettes $n_V$ in the RK-wave function for zero flux sectors. We show ED data for smaller size clusters ($N=36$ and $32$) and Monte Carlo data for larger (up to $N=576$ and $512$). The solid lines show the fit to the finite size corrections, Eq.~(\ref{eq:MC_intercept_finite_size}).
    \label{fig:MC_intercept}}
\end{figure}

For Monte Carlo calculations, we used clusters with $N=L^2$ geometry up to $N=576$ sites and $N=2L^2$ geometry up to $N=512$. The finite size dependence of the $n_V = \langle N_V \rangle / N$ and $n_{\text{II}} = \langle N_{\text{II}} \rangle / N$ densities is shown in Fig.~\ref{fig:MC_intercept}, together with ED data for the $N=32$ and $N=36$ sites to check the consistency of the data. The finite size fitting functions are as follows.
The density of the type-II vertices and flippable plaquettes for the $N=L^2$ class of clusters (red curves in Fig.~\ref{fig:MC_intercept})
\begin{subequations}
\label{eq:MC_intercept_finite_size}
\begin{align}
\langle n_{\text{II}} \rangle_{(0,0)} &=(0.61994 \pm 0.00005) - (0.56 \pm 0.02)  N^{-1}  \nonumber\\
&\phantom{=} + (2.3 \pm 0.7) N^{-2} \,,\\
\langle n_V \rangle_{(0,0)} &= (0.41579 \pm 0.00006) + (0.72 \pm 0.01) N^{-1} \nonumber\\
&\phantom{=} + (0.03 \pm 0.8) N^{-2} \,.
\end{align}
The same for the clusters with the $N=2 L^2$ geometry (green curves in Fig.~\ref{fig:MC_intercept})
\begin{align}
\langle n_{\text{II}} \rangle_{(0,0)} &= (0.61991 \pm 0.00007) - (0.55 \pm 0.02) N^{-1} \nonumber\\
&\phantom{=} - (2.3 \pm 1.2) N^{-2} \,,\\
\langle n_V \rangle_{(0,0)} &= (0.41582 \pm 0.00008) + (0.71 \pm 0.02) N^{-1} \nonumber\\
&\phantom{=} + (1.8 \pm 1.4) N^{-2} \,.
\end{align}
\end{subequations}
The expectation values for both cluster geometries extrapolate to the same values in the thermodynamic limit well within the error bars.

It is instructive to evaluate the ratio $\langle n_{\text{II}}\rangle/ \langle n_{\text{I}}\rangle$ in the RK wave function for different flux sectors. The density of the type-I vertices is $\langle n_{\text{I}}\rangle = 1-\langle n_{\text{II}}\rangle$, so
 for $\mathbf{m} = (0,0)$ we get
 \begin{equation}
 \frac{ \langle n_{\text{II}}\rangle_{(0,0)}}{ \langle n_{\text{I}}\rangle_{(0,0)}}
 = \frac{0.6199 \pm 0.0001}{0.3801 \pm 0.0001} = 1.6308 \pm 0.0006 \,.
 \label{eq:nIInIratio}
\end{equation}
Were the vertices uncorrelated, we would expect $ \langle n_{\text{II}}\rangle/ \langle n_{\text{I}} \rangle = 2$ instead of 1.63. This ratio improves as the flux increases, the limiting case being the isolated states with type-II vertices only.

\begin{figure}[tb!]
        \centering
   \includegraphics[width = 0.9 \columnwidth]{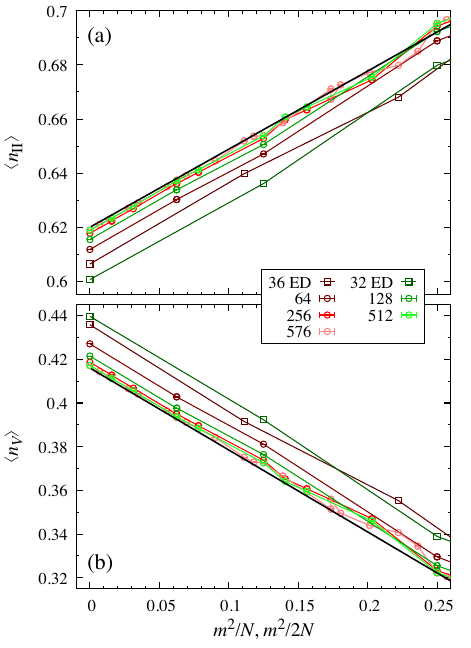}
    \caption{The expectation value of the densities of (a) type-II vertices $n_{\rm II}$ and (b) flippable plaquettes $n_V$ in the RK-wave function grows (decreases) linearly with the square of the total flux, $m^2=|\mathbf{m}|^2 = m_x^2 + m_y^2$. 
The evaluation is exact numerically for $N=32$ and 36, and we sampled the RK wave function by Monte Carlo for larger system sizes of up to 576 sites. For consistency, we divide $m_x^2 + m_y^2$ by $N$ for the $N=L^2$ clusters ($N=36,64,256,576$, denoted by red colors in the plots) and by $2N$ for the $N=2L^2$ size clusters ($N=32,128,512$, green in the plots). The black straight lines show the extrapolation to the thermodynamic limit.
    \label{fig:MC_1}}
\end{figure}

Inspired by the flux dependence of the energy in the $U(1)$ liquid in 3D \cite{Shannon2012, Pace_PRL_2021}, we plotted how the expectation values depend on the flux sectors for different cluster sizes in Fig.~\ref{fig:MC_1}. The plot reveals the linear dependence of $\langle n_{\text{II}}\rangle_{\mathbf{m}}$  on  $m^2/N$ for not too large values of the flux $\mathbf{m}$ and the slope appears to be independent of the size of the cluster. An additional factor of $2$ compensates for the geometry of the clusters.

\begin{figure}[bt!]
        \centering
\includegraphics[width = 0.9 \columnwidth]{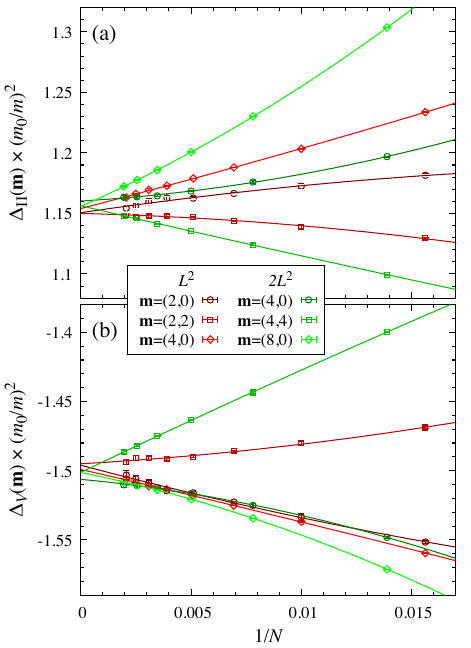}
    \caption{The finite size dependence of the gaps $\Delta_V(\mathbf{m}) = \langle N_V \rangle_{\mathbf{m}} - \langle N_V \rangle_{(0,0)}$ and  $\Delta_{\text{II}}(\mathbf{m}) = \langle N_{\text{II}} \rangle_{\mathbf{m}} - \langle N_{\text{II}} \rangle_{(0,0)}$ in a few selected flux sectors close to $\mathbf{m}=(0,0)$. The gaps collapse to a common value in the thermodynamic limit when divided by $m^2/m_0^2$. $m_0^2$ is the square of the unit flux, for the $L^2$ cluster $m_0^2=4$ and for the $2L^2$ cluster $m_0^2=8$. 
    \label{fig:MC_slope}}
\end{figure}

To further elucidate the linear dependence on $m^2$, we plotted the finite size dependence of the gaps
\begin{subequations}
\begin{align}
  \Delta_{\text{II}}(\mathbf{m}) &=
  \langle N_{\text{II}} \rangle_{\mathbf{m}} - \langle N_{\text{II}}\rangle_{(0,0)} \,,\\
    \Delta_{V}(\mathbf{m}) &=
    \langle N_{V} \rangle_{\mathbf{m}} - \langle N_{V}\rangle_{(0,0)}\;,
\end{align}
\end{subequations}
divided by $m^2/m_0^2$ in Fig.~\ref{fig:MC_slope}. The reason to divide $m^2$ by $m_0^2$ is the smallest nonzero flux ({\it i.e.}, unit of flux). It is $\mathbf{m}_0=(0,2)$ with $m_0^2=4$ in $N=L^2$ cluster with even $L$. In the $N=2L^2$ cluster, the unit of flux is $\mathbf{m}_0=(2,2)$ and, therefore, we divide $m^2$ by 8. The gaps in both cluster geometries are the same when threaded by the unit flux $\mathbf{m}_0$, independently of the system size. We collected the finite size scaled values of the $\Delta(\mathbf{m}) m_0^2/m^{2}$ in Tab.~\ref{tab:MC_slope}.
The numbers extracted from the lowest flux sectors are identical within the error bars, and we may conclude that
\begin{subequations}
\begin{align}
   \frac{m_0^2}{m^2} \Delta_{\text{II}}(\mathbf{m}) &= 1.154 \pm 0.004 \,, \\
   \frac{m_0^2}{m^2} \Delta_{V}(\mathbf{m}) &= - 1.500 \pm 0.004 \,.
\end{align}
\end{subequations}
Putting together with the $\mathbf{m} = (0,0)$ values in Eqs.~(\ref{eq:MC_intercept_finite_size}), we get the behavior of the expectation values in the thermodynamic limit
\begin{subequations}
\label{eq:ndensThermoLimit}
\begin{align}
  \langle n_{\text{II}} \rangle_{\mathbf{m}} &= (0.6199 \pm 0.0001) + (1.154 \pm 0.004) \frac{m^2}{m_0^2}\frac{1}{N} \,, \\
  \langle n_V \rangle_{\mathbf{m}} &= (0.4158 \pm 0.0001) - (1.500 \pm 0.004) \frac{m^2}{m_0^2}\frac{1}{N}  \,.
\end{align}
\end{subequations}
These are the solid black lines in Fig.~\ref{fig:MC_1}.

\begin{table}[bt]
\caption{The gaps $\Delta_{\text{II}}(\mathbf{m})$ divided by the square of the flux, $m^2$, for small values of $\mathbf{m}$ and clusters with two different geometries, $N=L^2$ ($m_0^2=4$) and $N=2L^2$ ($m_0^2=8$), in the thermodynamic limit. The finite size behavior is shown in Fig.~\ref{fig:MC_slope}.}
\label{tab:MC_slope}
\begin{ruledtabular}
\begin{tabular}{cccc}
geometry & $\mathbf{m}$ & $\Delta_{\text{II}}(\mathbf{m})\times \left(\frac{m_0}{m}\right)^{2}$ & $\Delta_V(\mathbf{m})\times \left(\frac{m_0}{m}\right)^{2}$ \\
\hline
$L^2$ & $(2,0)$ & $1.151 \pm 0.003$ & $-1.496 \pm 0.003$ \\
$L^2$ & $(2,2)$ & $1.150 \pm 0.002$ & $-1.495 \pm 0.002$ \\
$L^2$ & $(4,0)$ & $1.154 \pm 0.001$ & $-1.499 \pm 0.001$ \\
$2L^2$ & $(4,0)$ & $1.160 \pm 0.002$ & $-1.506 \pm 0.002$ \\
$2L^2$ & $(4,4)$ & $1.156 \pm 0.001$ & $-1.501 \pm 0.001$ \\
$2L^2$ & $(8,0)$ & $1.156 \pm 0.001$ & $-1.501 \pm 0.001$ \\
\end{tabular}
\end{ruledtabular}
\end{table}%

\begin{figure}[bh!]
        \centering
\includegraphics[width = 0.9 \columnwidth]{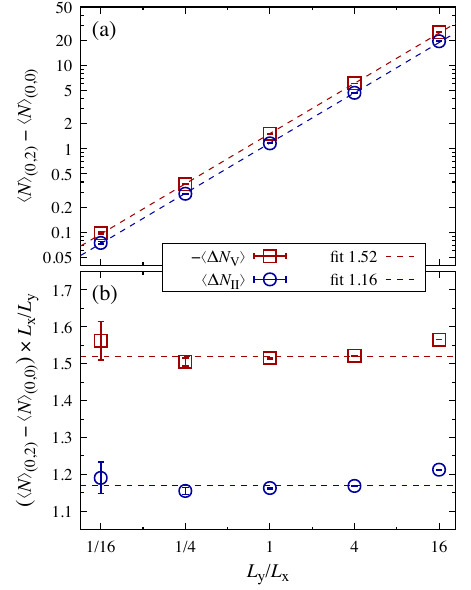}
    \caption{
The dependence of the gap of flippable plaquettes and type-II vertices between the smallest nonzero and the zero flux sector,
$\langle \Delta N_V \rangle = \langle \Delta N_V \rangle_{(0,2)} -  \langle \Delta N_V \rangle_{(0,0)}$
and
$\langle \Delta N_{\text{II}}\rangle = \langle \Delta N_{\text{II}} \rangle_{(0,2)} -  \langle \Delta N_{\text{II}} \rangle_{(0,0)}$
on the aspect ratio of the cluster. We calculated the  gap for $N = 256$ site rectangular clusters of different shapes: $L_x \times L_y = 4 \times 64, 8 \times 32, 16 \times 16, 32 \times 8$, and $64 \times 4$. We run 25 Monte Carlo simulations for each cluster with $5\times10^8$ steps each. We show $-\langle \Delta N_V \rangle$ for visual convenience. The large variation of the gap on the shape of the clusters is almost perfectly accounted for by the aspect ratio $L_y/L_x$ while keeping $L_x L_y$ constant: $\langle \Delta N_V \rangle \approx -1.52 L_y/L_x$ and $\langle \Delta N_{\text{II}} \rangle \approx 1.16 L_y/L_x$.
    \label{fig:aspect_ratio}}
\end{figure}

\subsection{The quantum electrodynamics of the RK wave function}
\label{sec:RK}

How do we understand the scaling of the expectation values at the RK point?
The local 2-in/2-out constraint at the vertices represents a divergence-free field. %
Associating the arrows with an electric field, one can build a kind of emergent quantum electrodynamics (QED) in the spin ice systems via the Gauss law, leading to a gapless $U(1)$ spin liquid phase. 
It has been widely studied in 3D models within the context of quantum spin ice \cite{Hermele2004,Banerjee2008, Shannon2012, Pace_PRL_2021}, and found that the $U(1)$ liquid extends beyond the RK point. 
In 2D, the gapless liquid phase is usually at the RK point only. All flux sectors have the same energy, and as we leave the RK point, a gap opens in the ground state flux sector, with possible exceptions \cite{Vishwanath2004, Fradkin2004, PhysRevLett.115.217202,Zhou2021}, as we will see later. The expectation values of the $n_V$ and $n_{\mathrm{II}}$ do not follow the behavior of the energy. While the energy values are degenerate, the $\langle n_V\rangle$ and $\langle n_{\mathrm{II}}\rangle$ values vary with the flux. 

The energy of the electric field is proportional to 
\begin{equation}
  \mathcal{E}_{\text{QED}} = \int_A \frac{1}{2} \varepsilon |\mathbf{E}|^2 dA
  \label{eq:EQED}
\end{equation}
where the integral is over area $A$. We neglect the ``magnetic" part of the emergent QED. 

In an effective theory, the average electric field on the lattice $\mathbf{E}$ is proportional to the $\mathbf{m}$.
Let us reverse a single arrow to see how the electric field emerges. 
It creates two vertices that are neither type-I nor type-II: a vertex with a 3-in/1-out (charge) and a vertex with a 1-in/3-out (anti-charge) arrows. 
These vertices can be considered as fractional charges, spinons, or monopoles, according to the actual physical problem we apply the Q6VM model. Moving one of these defects (charges) across the periodic boundary by reversing other arrows and eventually annihilating them makes a loop we considered e.g. in Fig.~\ref{fig:Clusters}. It changes the flux sector and introduces a finite electric field $\mathbf{E}$ when the arrows are coarse-grained. The strength of the average field is proportional to the density of the flux lines, $\mathbf{E}= q \mathbf{m}/L$ for the $N=L^2$ shape cluster and $\mathbf{E}= q \mathbf{m}/2L$ for the $N=2L^2$ shape cluster, where $q$ is the charge of the monopole. Taking the area $A=N$, squaring the $\mathbf{E}$ and replacing it into Eq.~(\ref{eq:EQED}), we get for the energy 
\begin{equation}
  \mathcal{E}_{\text{QED}} =  2 \varepsilon q^2 \frac{m^2}{m_0^2} \;.
  \label{eq:EQEDm2}
\end{equation}
At the RK point, the degeneracy implies $\varepsilon=0$. 

The expectation values of the 
$\langle n_{\text{II}} \rangle_{\mathbf{m}}$ 
and 
$\langle n_V \rangle_{\mathbf{m}}$ 
in Eq.~(\ref{eq:ndensThermoLimit}) are also quadratic function of the electric field. To further corroborate this statement, in Fig.~\ref{fig:aspect_ratio}, we show the dependence of the expectation values of $n_{\text{II}}$ and $n_V$ for a fixed number of sites but changing the aspect ratio of the rectangular cluster. The gap varies according to our expectations, the flux density.

Since, in the vicinity of the RK point, the energy follows Eq.~(\ref{eq:E_RK_1storder}), comparing the flux-dependent part with Eq.~(\ref{eq:EQEDm2}), we get
\begin{align}
   2 \varepsilon q^2 \frac{m^2}{m_0^2} &= (V-t) \langle N_V \rangle_{\mathbf{m}} + \mu \langle N_{\text{II}} \rangle_{\mathbf{m}} \nonumber \\
   &\approx \left[-1.500 (V-t)  + 1.154 \mu \right] \frac{m^2}{m_0^2}
\end{align}
that is
\begin{equation}
    \varepsilon \approx \frac{1}{2 q^2}\left[-1.500 (V-t)  + 1.154 \mu \right]\;.
    \label{eq:epsilon}
\end{equation}
For energies above the small gap in the $\mathbf{m} = (0,0)$ sector, we expect the system to follow the energy of the emergent QED with a tunable  $\varepsilon$ permittivity. This is in the spirit of Ref.~\cite{Pace_PRL_2021}, which considers the emergent QED in the 3D quantum spin-ice model.

\subsection{The phase boundaries emanating from the RK point}

\begin{figure}[tb!]
        \centering
   \includegraphics[width = 0.9 \columnwidth]{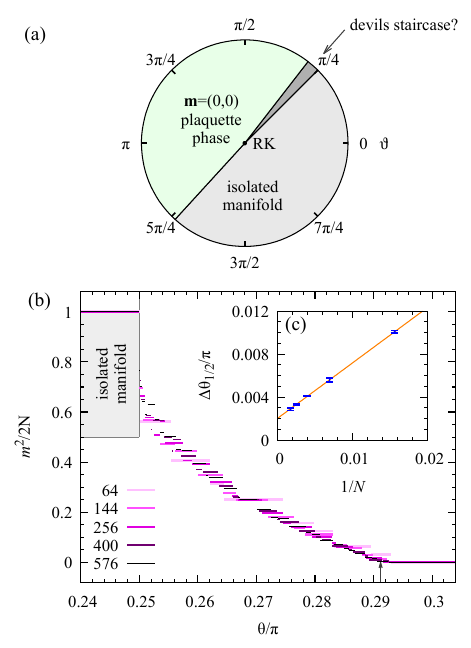}
    \caption{\label{fig:MC_top_secs_vs_theta}
    (a) The phase diagram around the $V=t$ and $\mu=0$ RK point. The parameter $\theta$ is defined by Eq.~(\ref{eq:vmuparametrization}). A first-order phase transition occurs between the isolated states and the plaquette phase at $\theta = 1.264 \pi$. For $\pi/4 < \theta < 0.291 \pi$, when  $\mu$ is positive, the two phases are separated by a region in which the flux sectors interpolate from the isolated manifold with maximal flux to the $\mathbf{m} = (0,0)$. The boundary $\theta = \pi/4$ (line $V = \mu + t $ for $\mu>0$) of the isolated states is exact and also holds away from the RK-point. The $\mathbf{m} = (0,0) $ flux sector is the ground state for $0.291 \pi <\theta <  1.264 \pi$.
    (b) We plot $m^2 = m_x^2 + m_y^2$ of the flux sectors with minimal energy as a function of the parameter $\theta$ in the vicinity of the RK-point  [$\delta \to 0$ in Eq.~(\ref{eq:vmuparametrization})] for $\pi/4 < \theta < 0.291 \pi$, where we expect the devil's staircase. The flux monotonically decreases with increasing $\vartheta$. The plot suggests a finite-width plateau at $m^2/2N=1/4$, corresponding to the flux sector $\mathbf{m}=(L/2, L/2)$. Calculations were done for up to $576$ sites in the $N=L^2$ type clusters. 
 (c) The finite size scaling of the width of the $\mathbf{m}=(L/2, L/2)$ plateau indicates a tiny but finite width $\Delta \theta_{1/2} = 0.002 \pi$ in the thermodynamic limit.}
\end{figure}

To describe the small perturbations around the RK point, we parametrize the $V-t$ and $\mu$ with the angle $\theta$
\begin{subequations}
\label{eq:vmuparametrization}
\begin{align}
 V -t  &= \delta \cos \theta \,, \\
  \mu &= \delta \sin \theta \,,
\end{align}
\end{subequations}
where $\delta$ is some small energy scale. Next, for a value of $\theta$ and system size, we calculate the energy in Eq.~(\ref{eq:E_RK_1storder})
\begin{equation}
    \frac{E_{\mathbf{m}}}{N} =
    \delta \left(\cos \theta \langle n_{V} \rangle_{\mathbf{m}}
    + \sin \theta\langle n_{\rm II} \rangle_{\mathbf{m}} \right)
    \label{eq:E_RK_1storder_delta}
\end{equation}
for each flux sector $\mathbf{m}$, and find for which it is minimal. The result of this energy minimization is presented in Fig.~\ref{fig:MC_top_secs_vs_theta}(a) for the full circle around the RK point. We recovered the phase boundaries anticipated from the ED calculations: the first-order phase transition between the isolated states and the plaquette phase for negative values of $\mu$ and the liquid phase for $\mu>0$.

To determine the boundary between the isolated states and the plaquette phase more precisely, we compare the energy density $\mu$ of the isolated states with the energy of the $\mathbf{m}=(0,0)$ sector using Eq.~(\ref{eq:E_RK_1storder}), that leads to the following equation: 
\begin{equation}
  \mu = (V-t) \langle n_{V} \rangle_{(0,0)} + \mu \langle n_{\text{II}} \rangle_{(0,0)} \,.
\end{equation}
For the angle $\theta_{\text{1st}} $  we then get
\begin{align}
\tan \theta_{\text{1st}}
  &= \frac{\langle n_V \rangle_{(0,0)}}{1-\langle n_{\text{II}}\rangle_{(0,0)}}
  = \frac{0.4158 \pm 0.0001}{0.3801 \pm 0.0001} \nonumber\\
  &= 1.0939 \pm 0.0004
\end{align}
 in the thermodynamic limit, using the extrapolations given in  Eqs.~(\ref{eq:MC_intercept_finite_size}), so $\theta_{\text{1st}} = (1.26427 \pm  0.00006) \pi$, taking into account that both $V-t$ and $\mu$ are negative at this boundary. Let us note that in the denominator the $1-\langle n_{\text{II}}\rangle_{(0,0)} = \langle n_{\text{I}}\rangle_{(0,0)}$, the density of the type-I vertices appears.

In Fig.~\ref{fig:MC_top_secs_vs_theta}(b), we zoom in on the tiny region where the liquid phase appears. The isolated manifold gives the ground state up the $\theta = \pi/4$, in full agreement with Eq.~(\ref{eq:isoBoundary}) in Sec.~\ref{sec:isolated}. The flux sector first appears next to isolated
manifold is the $\mathbf{m} = (L-2,L-2)$ in the clusters having $N=L^2$ sites.

To get the boundary between the plaquette and the liquid phases, we compare the energies of the $\mathbf{m}=(0,0)$ and small $\mathbf{m}$ flux sectors. This involve the gaps $\Delta_{V}(\mathbf{m})$ and $\Delta_{\text{II}}(\mathbf{m})$ and provide the 
\begin{equation}
  0 = (V-t) \Delta_{V}(\mathbf{m}) + \mu \Delta_{\text{II}}(\mathbf{m}) \,.
 \end{equation}
 condition so that
\begin{align}
  \tan \theta_{\text{2nd}}
 & = -\frac{\Delta_{V}(\mathbf{m})}{\Delta_{\text{II}}(\mathbf{m})} \nonumber\\
 &  = \frac{1.5 \pm 0.004}{ 1.154 \pm 0.004} =1.300 \pm 0.006
 \,,
 \end{align}
in the thermodynamic limit for values of $\mathbf{m}$ tabulated in Tab.~\ref{tab:MC_slope}. This translates to $\theta_{\text{2nd}} = (0.2912 \pm 0.0007 ) \pi$, indicated by the tiny arrow in Fig.~\ref{fig:MC_top_secs_vs_theta}(b). The window for the liquid state is thus tiny, about $\theta_{\text{2nd}} - \pi/4 \approx 0.041 \pi$. We note that the value of $\theta_{\text{2nd}}$ is where the permittivity $\varepsilon$ in Eq.~(\ref{eq:epsilon}) changes sign.

It is difficult to resolve the precise character of the liquid phase. It is unclear whether the topological sectors increase continuously or whether we are faced with an infinite sequence of incommensurate states, exemplifying a "devil's staircase" (also called "Cantor deconfinement") \cite{Vishwanath2004, Fradkin2004, PhysRevLett.115.217202,Zhou2021}. Possible evidence for the latter scenario is the plateau at half maximum flux, $\mathbf{m} = (L/2, L/2)$. Analysis of finite-size scaling suggests a finite width of the plateau, which is about 5\% of the size of the liquid phase, see Fig.~\ref{fig:MC_top_secs_vs_theta}(c). It is adjacent to the $\mathbf{m} = (L/2-2, L/2+2)$ and $\mathbf{m} = (L/2, L/2-2)$ flux sectors; however, this does not follow assuming a perfect $m^2$ dependence of the expectation values on the flux.

\section{Structure factors}
\label{sec:structure_factors}

\begin{figure*}[t!]
  \includegraphics[width = 0.99\textwidth]{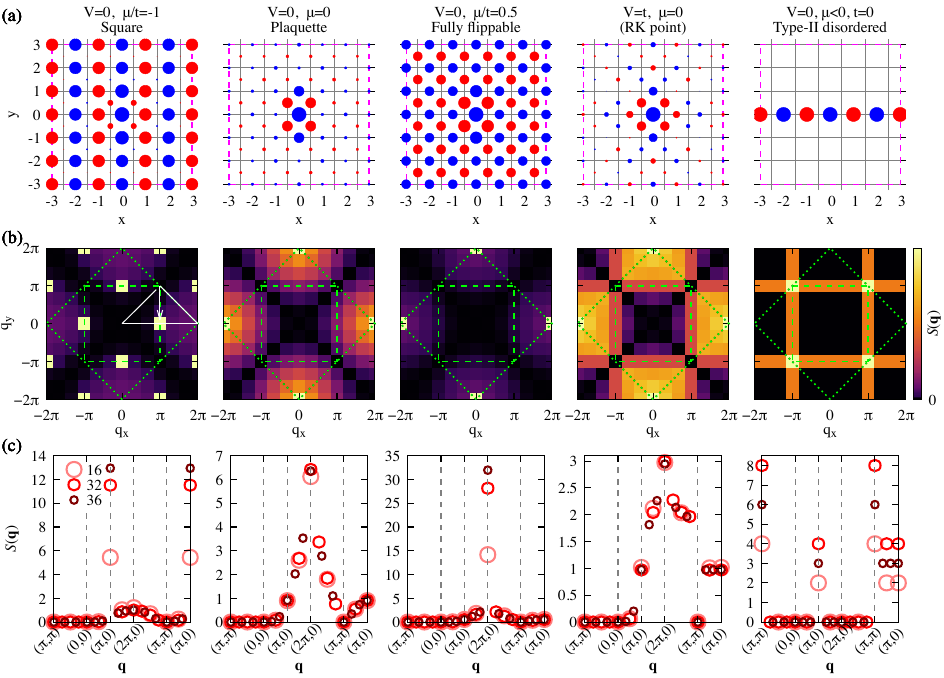}
  \caption{
  (a) The bond correlation function $ C^h(x,y) $ [Eq.~(\ref{eq:defCxyh})] in real space for the 36-site cluster (the coordinate of the horizontal bond at the center is $(\frac{1}{2},0)$ in the lattice, the thin grey lines stands for the lattice). It is equal to $C^{hh}(x,y)$ on the horizontal bonds, defined by Eq.~(\ref{eq:defCxyhh}), and to $C^{vh}(x+\frac{1}{2},y-\frac{1}{2})$ on the vertical bonds [see Eq.~(\ref{eq:defCxyvh})]. 
  From left to right, we present correlations in the square, plaquette, fully flippable phases, at the RK point, and in the classical disordered manifold.
  The area of the disks is proportional to the value of $S(x,y)$; blue indicates positive and red negative values.
  (b) The density plot of the structure factor $S(\mathbf{q})$. 
  The green dashed square encloses the Brillouin zone of the 36-site square cluster, while the green dotted line is the boundary of the extended Brillouin zone containing $2N=72$ $\mathbf{q}$ points. 
  (c) The $S(\mathbf{q})$ along the path $\mathbf{q} = (\pi,\pi) \to (0,0) \to (2\pi,0) \to (\pi,\pi) \to (\pi,0) $ drawn in white in the leftmost panel in (b). 
  The structure factor diverges with system size at $\mathbf{Q} = (\pi,0)$ in the square phase ($1^\text{st}$ column) and at $\mathbf{Q} = (2\pi,0)$ in the fully flippable phase ($3^\text{rd}$ column).
   The structure factor is diffuse in the plaquette phase, with a peak centered at $ \mathbf{Q}=(2\pi,0)$. At the RK point ($4^\text{th}$ column), the value of the structure factor strongly depends on the direction we approach the $(\pi,\pi)$,  $S(\pi+ \delta,\pi+\delta):S(\pi,\pi+\delta):S(\pi+ \delta,\pi-\delta)=0:1:2$ as $\delta \to 0$, demonstrating the non-analytic behavior of the pinch point.
   In the disordered manifold ($5^\text{th}$ column), we see subdivergent lines along $\mathbf{Q} = (\pm 2\pi, q)$ and $\mathbf{Q} = (q, \pm 2\pi)$. 
  \label{fig:SQ}
}
\end{figure*}

In this section, we determine the zero-temperature correlation functions and the structure factors using exact diagonalization. 
We will first discuss the structure factor  in the fully packed loop representation and then the magnetic structure factor in the arrow representation. 
The latter will allow us to compare our results to the ones observed in the artificial spin ice with superconducting flux qubits by King {\it et al.} in Ref.~\cite{King2021}.

\subsection{Correlations in fully packed loop representation}

 We define the correlation function with respect to a horizontal bond as 
 \label{eq:defCxy}
\begin{align} 
 C^h(x,y) &= \langle \text{GS} | n_{(x+\frac{1}{2},y)}n_{(\frac{1}{2},0)} | \text{GS} \rangle ,  \label{eq:defCxyh}
\end{align}
where $| \text{GS} \rangle$ is the translationally invariant ground state. $n_\mathbf{r}$ measures whether the bond centered at $\mathbf{r}$ is occupied ($n_\mathbf{r} = 1$) or not ($n_\mathbf{r} = -1$).
Since the $| \text{GS} \rangle$ transforms according to the trivial irreducible representation, the $\langle \text{GS} | n_{\mathbf{r}} | \text{GS} \rangle = 0$ and we can use the above definition of $C^h(x,y)$.
The $x$ and $y$ values are either both integers or half-odd integers. We calculate the ground state wave function $| \text{GS} \rangle$ using the L\'anczos algorithm in $N=16$, $32$, and $36$ site clusters with periodic boundary conditions for a few selected parameters, representing the different phases. The exact diagonalization provides a fully symmetric $| \text{GS} \rangle$ in a finite cluster with periodic boundary conditions (the $|\text{GS}\rangle$ is in the $(0,0)_e$ symmetry sector, see Figs.~\ref{fig:GUS32} and \ref{fig:GUS36}).
For practical purposes, we introduce the
\begin{subequations}
 \label{eq:defCxy}
\begin{align} 
 C^{hh}(i_x,i_y) &= \langle \text{GS} | n_{(i_x+\frac{1}{2},i_y)}n_{(\frac{1}{2},0)} | \text{GS} \rangle ,  \label{eq:defCxyhh}
\\
 C^{vh}(i_x,i_y) &= \langle \text{GS} | n_{(i_x,i_y+\frac{1}{2})}n_{(\frac{1}{2},0)} | \text{GS} \rangle ,  \label{eq:defCxyvh}
\\
C^{hv}(i_x,i_y) &= \langle \text{GS} | n_{(i_x-\frac{1}{2},i_y)}n_{(0,-\frac{1}{2})} | \text{GS} \rangle ,  \label{eq:defCxyhv}
\\
 C^{vv}(i_x,i_y) &= \langle \text{GS} | n_{(i_x,i_y-\frac{1}{2})}n_{(0,-\frac{1}{2})} | \text{GS} \rangle , \label{eq:defCxyvv}
\end{align}
\end{subequations}
where the coordinates $(i_x,i_y)$ are integers.
 The $C^{hh}(i_x,i_y)$ is a correlation function between horizontal bonds, $C^{vv}(i_x,i_y)$ between vertical bonds, and $C^{vh}(i_x,i_y)$ and $C^{hv}(i_x,i_y)$ between orthogonal bonds. They provide sufficient information to obtain both the density and the magnetic correlation function in real and reciprocal space. 
How they behave under the action of the point group symmetries is described in Appendix~\ref{sec:app_hsym}.

Fig.~\ref{fig:SQ}(a) displays the bond-bond correlation function in the ordered square,
plaquette,
and fully flippable phase,
as well as for the quantum-disordered RK point from ED calculations on the 36-site cluster, and for the disordered phase boundary in the classical phase diagram.
%
While the bond-bond correlations decay rapidly in the plaquette phase and at the RK point, 
the long-range pattern of the ordered loops manifests itself in the classical square and fully flippable phase. In the square phase, when the central horizontal bond at $(i_x,i_y)=(0,0)$ in Fig.~\ref{fig:SQ}(a) is occupied (blue disk), all the horizontal bonds in the same columns are also occupied and the next column of horizontal bonds is empty (red disks), in full accordance with Fig.~\ref{fig:Squares}(b). For the fully flippable phase, the occupied horizontal bond invokes the occupation of the horizontal bonds (all the horizontal bonds are blue, and all the vertical bonds are red), c.f. Fig.~\ref{fig:FFs}(b). In the disordered manifold of the classical Rys-F model (the boundary between the isolated and square phase in Fig.~\ref{fig:CPD}), the bond-bond correlations are finite only along a line (otherwise, the average over the disorder nulls the correlations). 

The Fourier transform of the real space correlation function is the structure factor
\begin{equation}
  S(\mathbf{q}) = \frac{1}{2N} \sum_{\mathbf{r},\mathbf{r'}} e^{i \mathbf{q}\cdot(\mathbf{r}-\mathbf{r'})} 
  \langle \text{GS} | n_{\mathbf{r}} n_{\mathbf{r'}} | \text{GS} \rangle \;,
  \label{eq:defSq}
\end{equation}
where both $\mathbf{r}$ and  $\mathbf{r'}$ in the sum run over the $N$ horizontal and $N$ vertical bonds.
Separating the vertical and horizontal bonds, we arrive to the 
\begin{multline}
  S(\mathbf{q}) =
 \frac{1}{2} \sum_{i_x,i_y} e^{i (i_x,i_y) \cdot  \mathbf{q}} 
 \left[ 
    C^{hh}(i_x,i_y)   + C^{vv}(i_x,i_y) \right.
\\
 \left. + C^{vh}(i_x,i_y)  e^{i (-\frac{1}{2},\frac{1}{2})\cdot\mathbf{q}} 
  + C^{hv}(i_x,i_y)  e^{i (-\frac{1}{2},\frac{1}{2})\cdot\mathbf{q}}  
    \right]
\end{multline}
expression. Using the symmetry properties described by Eqs.~(\ref{eq:Chhsym})-(\ref{eq:CvhChvsym}), one can show that the  $S(\mathbf{q})$ is real and satisfies the full $D_4$ point group symmetry in the $\mathbf{q}$ space.

Since the centers of the horizontal and vertical bonds form a square lattice rotated by 45$^\circ$ and lattice constant $1/\sqrt{2}$, the $S(\mathbf{q})$ is periodic in the reciprocal space for momenta shifts by $(2\pi,2\pi)$ and  $(2\pi,-2\pi)$:   $S(\mathbf{q}) = S(\mathbf{q}+(2\pi,2\pi)) $ and $S(\mathbf{q}) = S(\mathbf{q}+(2\pi,-2\pi))$. Therefore the first and second Brillouin zones together contain all the relevant information, and we will call their union the extended Brillouin zone (EBZ). 
The structure factor satisfies the 
\begin{equation}
   \sum_{\mathbf{q} \in \text{EBZ}} S(\mathbf{q}) = 2N
   \label{eq:sumrule}
\end{equation}
sum rule, where the sum is over the $2N$ $\mathbf{q}$-points in the extended Brillouin zone. 
 Fig.~\ref{fig:SQ}(b) displays the structure factor for $-2\pi \leq q_x \leq 2\pi$ and $-2\pi \leq q_y \leq 2\pi$. To demonstrate finite-size effects, we plot $S(\mathbf{q})$ for three system sizes, $N=16$, $32$, and $36$ in Fig.~\ref{fig:SQ}(c) along a path in the reciprocal space drawn in the leftmost panel of Fig.~\ref{fig:SQ}(b).  

Before discussing the structure factors in detail, let us note a general feature present in all the plots: the weight disappears along the $\mathbf{q} = (q,q)$ and $(q,-q)$ lines in the reciprocal space. The vanishing weight is the consequence of the ice rule, which imposes the local divergence-free constraint for the allowed configurations~\cite{Youngblood1981}. In the fully packed loop representation, the ice rule  manifests itself as 
\begin{equation}
  n_{(i_x+\frac{1}{2},i_y)} + n_{(i_x,i_y+\frac{1}{2})} + n_{(i_x-\frac{1}{2},i_y)} + n_{(i_x,i_y-\frac{1}{2})} = 0,
\end{equation}
 i.e., the sum of occupations of bonds sharing the same vertex shall vanish. This implies that for an arbitrary configuration, the sum of the bond occupations along a diagonal, $\sum_i (-1)^j n_{(j+i/2, 1/2 \pm i/2)}$ is a constant, where the $\pm$ determines the orientation of the diagonal. Therefore, the Fourier transform along the $\mathbf{q} = (q,\pm q)$ vanishes except at $\mathbf{q} = (\pi,\pm \pi)$.
  The weight at $(\pi,\pm \pi)$ is related to the flux sector, and one finds that $S(\pi,\pm \pi)= (m_x\pm m_y)^2/2$.

Let us consider the ordered states. For the fully flippable phase, diverging Bragg peaks appear at the ordering wave vectors are $\mathbf{Q} = (0,2\pi)$ and  $\mathbf{Q} = (2\pi,0)$. These points are the $\Gamma$ points in the second Brillouin zone and reflect the fact that the fully flippable state does not break translational symmetry. The positions and amplitudes of the peaks agree with Eq.~(\ref{eq:SQ_FF_calc}) we got from an analytical calculation presented in Appendix~\ref{sec:SQ_FF_calc}.

The square phase shows a Bragg-peak at the ordering vectors $\mathbf{Q} = (0,\pm\pi)$ and at $\mathbf{Q} = (\pm\pi,0)$, in full consistency with the analytical calculation presented in Appendix~\ref{sec:SQ_Sq_calc} and summarized by Eq.~(\ref{sec:SQ_Sq_calc}). The structure factor at these ordering wave vectors diverges with the system size, as demonstrated in Fig.~\ref{fig:SQ}(c), where the $S(\mathbf{Q})$ doubles between the 16- and 32-site results. 

Though ordered, the plaquette phase has no Bragg peaks. The structure factor peaks at $\mathbf{Q} = (0,2\pi)$ and $(2\pi,0)$. The peaks do not diverge but only depend weakly on the system size, with a diffuse scattering visible around them. We calculated the shape of the diffuse scattering using the variational wave function in Appendix~\ref{sec:SQ_Plaq_calc}, and compared it to the numerical calculation in Fig.~\ref{fig:SQPlaq}. 

The diffuse scattering changes shape at the RK point, where pinch-point singularities appear at $\mathbf{q} = (\pi,\pm\pi)$. At those points, the value of the structure factor is not an analytic function of the momenta and is of the form
\begin{equation}
  S(\pi+ k_x,\pi+ k_y) \propto 1 -  \frac{2 k_x k_y}{k_x^2 + k_y^2}
\end{equation}
in the vicinity of the $\mathbf{q} = (\pi,\pi)$, where the $k_x$ and $k_y$ are small \cite{Youngblood1981}.

The structure factor disordered manifold for the $V=t=0$, $\mu<0$ shows sub-divergent lines in the Brillouin zone, the lines are at $\mathbf{q} = (\pi z, q')$ and $(q',\pi z)$, where $z\in \mathbb{Z}$ is an integer and the coordinate $q'$ runs over all possible values for the corresponding momentum (see Appendix~\ref{sec:SQ_Dirord_calc} for the exact analytical treatment).

\subsection{Magnetic structure factor}
\label{sec:experiments}

\begin{figure*}[t!]
  \includegraphics[width = 0.99\textwidth]{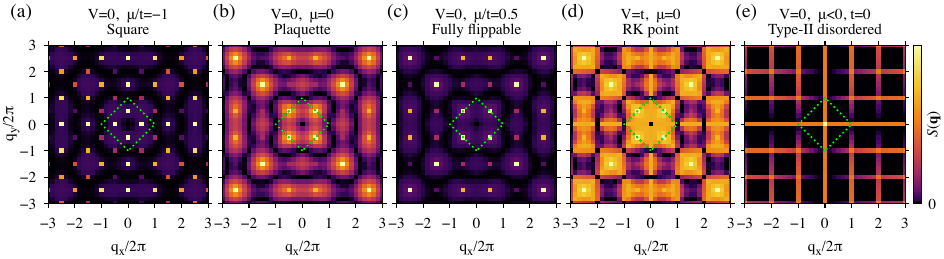}
  \caption{  \label{fig:SQneutron}
  The magnetic structure factor $\tilde{S}(\mathbf{q})$ in the arrow representation, given by Eq.~(\ref{eq:defSqneutron}) and calculated by exact diagonalization on the 36-site cluster for (a)-(d) and analytically for (e). The ordered phases are (a) the square phase for $\mu/t = -1$, (b) the plaquette phase for $\mu=0$, and (c) the fully flippable for $\mu/t=0.5$, keeping $V=0$ in all three cases. We find divergent Bragg peaks in the square and the fully flippable phases, while $\tilde{S}(\mathbf{q})$ remains diffuse in the plaquette phase. (d) $\tilde{S}(\mathbf{q})$  at the RK-point ($V=t$ and $\mu=0$) displays the pinch points at $\mathbf{Q} = (2 \pi z_x,2 \pi z_y)$, where $z_x,z_y\in\mathbb{Z}$. (e) The magnetic structure factor of the $V=t=0$ classical disordered manifold. The amplitudes of the sub-divergent horizontal and vertical lines are woked out in  Appendix~\ref{sec:SQ_Dirord_calc}. 
  The green dotted square denotes the boundary of the extended Brillouin zone, the same as in Fig.~\ref{fig:SQ}(b).
}
\end{figure*}

The magnetic structure factor in spin systems is measured by neutron scattering. The neutron cross-section is proportional to  
\begin{multline}
  \tilde{S}(\mathbf{q}) =
  \sum_{\alpha,\beta}  
  \left(\delta_{\alpha,\beta}-\frac{q_\alpha q_\beta}{q^2}\right) 
 \\ \times  \frac{1}{2N} \sum_{\mathbf{r},\mathbf{r'}} 
  \langle \text{GS} | M^{\alpha}_{\mathbf{r}} M^{\beta}_{\mathbf{r'}} | \text{GS} \rangle  
  e^{i (\mathbf{r}-\mathbf{r'})\cdot\mathbf{q}}\;,
  \label{eq:defSqneutron}
\end{multline}
where $M^{\alpha}_{\mathbf{r}}$ is the $\alpha$-component of the spin operator at site $\mathbf{r}$. It is customary to calculate the structure factor above for the six vertex models as well. 
In the two-dimensional case, we choose arrows instead of spins so that $M^x=\pm 1$ and $M^y=0$ for a horizontal arrow on a horizontal bond and $M^x=0$ and $M^y=\pm 1$ for a vertical arrow on a vertical bond, following Eq.~(\ref{eq:FPL2Arrows}).
Furthermore, Eq.~(\ref{eq:FPL2Arrows}) allows us to express the $\tilde{S}(\mathbf{q})$ using the correlation function we defined in Eqs.~(\ref{eq:defCxy}). 
Eventually, we arrive to 
\begin{multline}
  \tilde{S}(\mathbf{q}) =
 \frac{1}{2} \sum_{i_x,i_y} e^{i (i_x,i_y) \cdot  \left[\mathbf{q}-(\pi,\pi)\right]} 
 \\
 \times\left\{
  \frac{q_y^2}{q^2} C^{hh}(i_x,i_y)  
 + 
  \frac{q_x^2}{q^2} C^{vv}(i_x,i_y) 
 \right.
 \\ 
 \left.
 - \frac{q_x q_y}{q^2} 
   \left[ C^{vh}(i_x,i_y)  +  C^{hv}(i_x,i_y) \right]
  e^{i (-\frac{1}{2},\frac{1}{2})\cdot\mathbf{q}}
 \right\}.
\end{multline}
Figure \ref{fig:SQneutron} displays the spin structure factor $\tilde{S}(\mathbf{q})$ for the typical examples shown in Fig.~\ref{fig:SQ}(b). As for $S(\mathbf{q})$, Bragg peaks dominate the spin structure factor in the square and the fully flippable phase, and diffuse scattering is seen for the plaquette phase and the RK-point. In the case of the RK point, the pinch points move to $\mathbf{Q}$ values that are now multiples of $2\pi$. 
The sub-divergent lines of the disordered manifold get a momentum-dependent intensity from the dipolar factor in Eq.~(\ref{eq:defSqneutron}), in addition to the momentum shift.

\subsection{Comparison with the structure factors in the artificial quantum 6-vertex model}

We can now compare our zero-temperature results for the structure factors with those observed in the artificial quantum spin-ice experiment. Our magnetic structure factor $\tilde{S}(\mathbf{q})$ should correspond to that in the "strong coupling" regime in Fig.~2 in Ref.~\onlinecite{King2021} ($J=J_{\text{MAX}}$ in their notation), where the frequencies (densities) of type-III and IV vertices are negligible. We shall bear in mind that the experiment dealt with a system with open boundary conditions at finite temperature while we work with periodic boundary conditions at zero temperature.

Let us start with the topmost row in Fig.~2 in Ref.~\onlinecite{King2021}, the "degenerate ice" with $J_\perp/J_\|=1$. At this point, the observed $N_{\text{I}}: N_{\text{II}} = 1:2$ ratio between the number of type-I and II vertices corresponds to their statistical weights, as the local Hilbert space on a site consists of two type-I and four type-II vertices, so neither of the vertices is favored in this case. The same ratio also occurs at the RK-point, where the ground state wave function is an equal-amplitude superposition of the allowed vertex configurations. We find excellent agreement with our Fig.~\ref{fig:SQneutron}(d) regarding the overall distribution of the weight. However, there is one notable difference: the pinch points at the $\mathbf{Q} = (2n_x \pi, 2 n_y \pi)$ show no weights in our Fig.~\ref{fig:SQneutron}(d), while in Ref.~\onlinecite{King2021} they are finite. 
A possible resolution of this discrepancy is that the weight at the center of the pinch point comes from finite flux sectors. 
The $\langle n_{\text{II}}\rangle_{(0,0)}/ \langle n_{\text{I}}\rangle_{(0,0)} \approx 1.631$ in Eq.~(\ref{eq:nIInIratio}) also supports this idea, to recover the measured $\langle n_{\text{II}}\rangle/\langle n_{\text{I}}\rangle \approx 2$ we shall take into account other flux sectors as well.
However, it is also possible that the temperature was so high that it concealed the quantum fluctuations and the complete disorder is a consequence of the almost identical Boltzmann-weights of ice rule configurations. Ref.~\onlinecite{Rougemaille2021} studied this case with classical Monte Carlo and noted the same structure factor.
%

Let us proceed to the second row in Fig.~2 in Ref.~\onlinecite{King2021}, denoted by ``Type-I bias". The choice of parameters resulted in slightly more type-I vertices than type-II vertices, with a ratio $N_{\text{I}}: N_{\text{II}} \approx 0.54:0.46$. According to our ED calculations shown in Figs.~\ref{fig:GUS32}(a) and \ref{fig:GUS36}(a), the experimental $N_{\text{I}}: N_{\text{II}} $ ratio corresponds to the $\mu/t \approx 0.12$ in the plaquette phase when $V=0$, close to the fully flippable boundary at $\mu/t \approx 0.29$  (we find $\langle N_{\text{I}} \rangle : \langle N_{\text{II}} \rangle = 0.6:0.4$ in our calculations at the phase boundary). 
In the experimental structure factor, we can identify both the Bragg peaks of the fully flippable phase with a diffuse scattering around, just like in our Fig.~\ref{fig:SQneutron}(c), and the finite rhombi-like structures at $(q_x,q_y) = (0,\pm 3\pi)$ and $(\pm3\pi,0)$ typical for the plaquette phase shown in Fig.~\ref{fig:SQneutron}(b). Therefore, it is hard to identify the ground state unambiguously. 
While the fully flippable phase is classical, the emergence of the diffusive structures compatible with the plaquette phase may indicate quantum effects at play.

Lastly, we turn to the third row in Fig.~2 in Ref.~\onlinecite{King2021}, which shows the case where $N_{\text{I}}: N_{\text{II}} \approx 0.19:0.81$ in the strong-coupling limit ("Type-II bias" with $J_\perp/J_\|=0.98$). We can reproduce the reported ratio of occupancies setting $\mu/t \approx -0.9$ in our $V=0$ exact diagonalization calculation, which is in the square phase but close to the plaquette phase. The magnetic structure factors are quite different: our calculation reveals apparent Bragg peaks with some diffuse scattering, while the experimental plot shows strong intensities along straight lines. But such structures appear in the finite temperature plots in Fig.~3(b) in Ref.~\onlinecite{Rougemaille2021} and our calculation of the magnetic structure factor in the $V=t=0$ disordered manifold. It implies that the thermal fluctuations destroy the quantum-mechanical order, and the observed state is a mixture of different flux sectors. 
The fact that quantum features are revealed in some cases and not in others may be related to the excitation energies in different phases, and how they compare to the temperature. However, this is only a hypothesis at this stage and would require further study.

Above, we compared the magnetic structure factors ${\tilde S}(\mathbf{q})$. The equivalent of the $S(\mathbf{q})$, defined by Eq.~(\ref{eq:defSq}), is shown in Fig.~S4 in the Appendix of Ref.~\onlinecite{King2021} (we note that the coordinate axes in Fig.~S4 are rotated by 45$^\circ$ in contrast to the magnetic structure factors ${\tilde S}(\mathbf{q})$ and the indicated Brillouin zone corresponds to our extended Brillouin zone.).

\section{Summary}
\label{sec:summary}

We studied the ground state properties of a quantum six vertex model on the square lattice that distinguishes the type-I and type-II vertices. We established the zero-temperature phase diagram and the static correlation functions in real and momentum space using analytical and numerical methods.

Regarding the classical ($t=0$) model, we found three extended phases in the $\mu$--$V$ parameter space. As discussed in sections \ref{sec:classical}-\ref{sec:CPD}, the twofold degenerate fully flippable phase contains only type-I vertices. The fourfold degenerate square phase and the sub-extensive manifold of isolated configurations consist of only type-II vertices. Figure~\ref{fig:CPD} summarizes the classical phase diagram.
The fully flippable phase is the analog of the ``antiferroelectric'' phase in the Rys $F$ model~\cite{Rys1963}, while the subextensive boundary between the square phase and the isolated manifold is equivalent to the ``disordered'' phase identified in Refs.~\cite{Lieb1967, Sutherland1967_PhysRevLett.19.103}. 

Classification of the configurations by the number of flippable plaquettes and type-II vertices revealed that the two numbers are correlated and form a triangle presented in Fig.~\ref{fig:nVnII_map}. 
Configurations at the corners of the triangle define the three classical phases mentioned above.  
In Sec.~\ref{sec:Symmetries}, we identified the symmetries broken in these phases, constructed diagonal order parameters as irreducible representations of the $\mathsf{D_4}$ point group using vertices, and wrote down the Landau free energy for the model taking into account the charge conjugation symmetry.

To derive the phase diagram for the quantum model, we applied the L\'anczos method to diagonalize the model on finite-size clusters ($N \leq 36$) with periodic boundary conditions. We calculated the expectation values of the order parameters and followed the level crossings in the low-energy spectra in a wide range of parameters (see Figs.~\ref{fig:GUS32} and \ref{fig:GUS36} for $V=0$ case). 
Figure~\ref{fig:ED_phase_diag} summarizes our phase diagram.
Interestingly, the boundary of the plaquette phase does not confine to small $V/t$ and $\mu/t$ but extends along the line $V= \mu + t$ for large values, together 
with a gapless liquid phase emanating from the quantum critical Rokhsar-Kivelson point. The extension of the quantum behavior results from the highly degenerate $V=\mu$ classical boundary so that $t$ competes with $V-\mu$.
A rapidly converging perturbation expansion in $t$ up to the fourth order confirmed the accuracy of the phase boundaries. 
Beyond the numerical methods, we applied Gerschgorin's theorem to obtain the exact boundary between the isolated and the liquid phases in Sec.~\ref{sec:isolated}. 
A simple variational treatment indicated the existence of a tricritical point between the square and plaquette phases, allowed by the Landau-free energy expansion.

We applied the Hellman-Feynman theorem to reveal the nature of the liquid phase and the splitting of the ground state degeneracy of the multicritical Rokhsar-Kivelson point in Sec.~\ref{sec:rainbow}.
Using Monte Carlo simulations for clusters up to 576 sites, we calculated the expectation values of the densities of flippable plaquettes and the type-II vertices scanning through the flux sectors. 
We obtained the phase boundaries emanating from the quantum critical point. The liquid phase unveils itself as a possible manifestation of the ``devil's staircase'' \cite{Vishwanath2004, Fradkin2004}, with evidence for a finite-width plateau at half the maximum flux sector.
The results also allowed us to study emergent quantum electrodynamics and to determine the electrical permittivity near the RK point.

Finally, in Sec.~\ref{sec:structure_factors}, we presented the zero-temperature structure factors for the various ordered phases and the RK point. We compared our results with the experiment on the artificial spin-ice system formed by superconducting qubits of Ref.~\cite{King2021}. While many features of our calculation and the measurement agreed, there were also some that we could not interpret using our zero-temperature calculation.

\acknowledgments

The authors acknowledge discussions with Nic Shannon, Hosho Katsura, and R. Ganesh. We acknowledge the financial support by the Hungarian NKFIH Grant Nos. K124176 and K142652. We wrote our code in Julia, an open-source programming language~\cite{Julia}.

\appendix

\section{Flux sectors}
\label{sec:Appendix_flux_sectors}

\begin{figure}[bh]
        \includegraphics[width = 0.8\columnwidth]{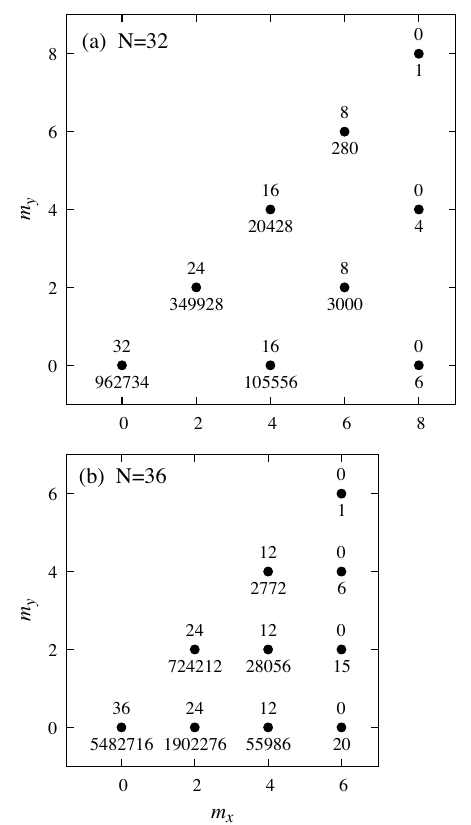}
    \caption{The flux sectors, their degeneracy (below the point), and max value of $N_V$ (above the point) for the (a) 32 site and (b) $N=36$ site clusters with PBC. For visualization purposes, only one irreducible octant is shown. The full diagram follows from symmetry.
    \label{fig:topsec_32_36}}
\end{figure}

Figure \ref{fig:topsec_32_36} shows the flux sectors in the 32- and 36-site cluster, together with the degeneracy and the maximal number of flippable plaquettes. The binomial coefficients give the number of isolated configurations in different flux sectors, as described in Sec.~\ref{sec:isolated}. We compare the number of configurations in the $(0,0)$ flux sector with the extensive degeneracy 
$W\propto (4/3)^{3N/2}$ in Table~\ref{tab:degs}.

\begin{table}[tb]
\caption{The degeneracy (deg.) of the $(0,0)$ flux sector for different system sizes $N$.  The third column is the calculation based on Lieb's formula, and the ratio with the actual degeneracy. The fifth and sixth columns are the estimate from Pauling's formula and the ratio. While the ratio is stable for Lieb's estimate, the Pauling formula underestimates the number of configurations.}
\label{tab:degs}
\begin{ruledtabular}
\begin{tabular}{rrrlrl}
   	&		& \multicolumn{2}{c}{Lieb} & \multicolumn{2}{c}{Pauling} \\
$N$	&deg.	& $(4/3)^{3N/2}$ & ratio & $(3/2)^{N}$ & ratio\\
\hline
$16$&$	990			$&$996.6			$&$1.0067$&$656.8$&$0.663$\\
$32$&$	962\,734	$&$993\,251.8	$&$1.0317$&$431\,439.9$&$0.445$\\
$36$&$	5\,482\,716	$&$5\,580\,739.8	$&$1.0179$&$2\,184\,164.4$&$0.398$\\
\end{tabular}
\end{ruledtabular}
\end{table}

\section{Inequalities for the 6VM on the square lattice}
\label{sec:inequalities}

\begin{figure}[tb]
    \centering
    \includegraphics[width=.8\columnwidth]{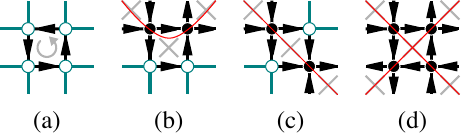}
    \caption{ 
The four symmetrically inequivalent vertex configurations at the corners of a plaquette.
(a) If the plaquette is flippable, the vertices are undetermined (open circles). 
(b) and (c) A non-flippable plaquette has at least two or (d) four type-II vertices (closed circles). The remaining vertices are undetermined (open circles). The red lines are the Faraday loops passing through the non-flippable vertices.\label{fig:Constraint3}}
\end{figure}

To determine the classical phase diagram, it is essential to know what values of the flippable plaquettes ($N_{V}$) and the type-II vertices ($N_{\rm II}$) characterize the 6-vertex configurations on the square lattice. Below, we consider finite systems with periodic boundary conditions containing $N$ vertices ($N$ plaquettes) and derive the inequalities (\ref{eq:GeomConstrains}).

$N_{\rm II} \leq N$: There cannot be more type-II vertices than there are vertices. Isolated and square configurations realize the case of equality, so the estimate is sharp.

$ 2 N_{V} + N_{\rm II} \leq 2 N$: 
Out of four plaquettes around a type-II vertex, at least two are non-flippable.
Let us fix $N_{\rm II}$ and find a configuration where the number of non-flippable plaquettes $N-N_{V}$ is minimal. The best strategy is to densely pack type-II vertices so that four type-II vertices block the same plaquette, just like in the square configurations. Detaching a type-II vertex would generate two new flippable plaquettes. Consequently, we get a lower bound for the number of non-flippable plaquettes: $2 N_{\rm II} / 4 \leq N-N_{\rm V}$. From this, Eq.~(\ref{eq:GeomConstrainsB}) follows.

$N \leq N_{V} + N_{\rm II}$:  Any non-flippable plaquette has at least two type-II vertices; in a specific case, it has four [solid circles in Fig.~\ref{fig:Constraint3}(b)-(d)].
 The vertex type of the open circles is not determined; they can be type-I or II.
We can construct loops of non-flippable plaquettes by connecting them via the type-II vertices denoted by the solid circles in the case of (b) and (c), or the loops intersect (d). 
These loops must close in a finite system and contain the same number of type-II vertices and non-flippable plaquettes. So if there are non-intersecting loops only, $N-N_V \leq N_{\text{II}}$ because there might be unaccounted type-II vertices on empty circles. It is equivalent to Eq.~(\ref{eq:GeomConstrainsC}). The intersection of the loops does not violate the inequality.

We recall that boundary conditions fundamentally influence the above inequalities. If we have open boundary conditions or infinite system size, it may not be true that the number of the non-flippable plaquettes is $N-N_{V}$, even the number of the plaquettes and the vertices might be different.

\section{A simple variational wave function }
\label{sec:variational_WF}

To describe the symmetry breaking of the plaquette state down to the square phase, we devise a simple variational wave function that interpolates between one of the plaquette [Eq.~(\ref{eq:Plaq_WFAD})] states and two square states as
\begin{align}
  | \Psi(p) \rangle &= \frac{1}{2+2p^2}
  \left[
  (1 + p) \mid \circlearrowleft \rangle_{\rm A}
  + (1 - p) \mid \circlearrowright \rangle_{\rm A}
  \right]
  \nonumber\\ & \phantom{=}
  \otimes  \left[
  (1 - p) \mid \circlearrowleft \rangle_{\rm D}
  + (1 + p) \mid \circlearrowright \rangle_{\rm D}
  \right]  \;.
  \label{eq:Plaq2Sq}
\end{align}
For $p=0$, it gives back the $| \Psi(0)  \rangle =| \text{PlAD} \rangle$, and for $p=\pm1$, it results in the classical square states 
$| \Psi(1)  \rangle = | \text{SqA} \rangle$ and $| \Psi(-1)  \rangle= | \text{SqD} \rangle$. 
The parameter $p$ is directly linked to the order parameter of the square phase,
\begin{equation}
   \langle \Psi(p) | \mathbf{O}_{\text{Sq}} | \Psi(p) \rangle = \frac{\sqrt{2}p}{1+p^{2}}\binom{1}{-1},
\end{equation}
while the plaquette order parameter is an even function of $p$ and is nonzero all the time,
\begin{equation}
   \langle \Psi(p) | O_{\text{Plaq}} | \Psi(p) \rangle = \frac{1 + 6 p^2 + p^4}{2 \left(1 + p^2 \right)^2} \approx \frac{1}{2} \left(1 + 4p^2 + \cdots \right) \;.
\end{equation}
We may compare these values to Tab.~\ref{tab:ordpar_values}.
The energy density $E(p)/N = \langle \Psi(p) | \mathcal{H} | \Psi(p) \rangle/N$ reads
 \begin{align}
   \frac{E(p)}{N} 
    & = \frac{V}{2} \left(1 + \frac{2 (1+p)^4 (1-p)^4}{(2+2p^2)^4}\right)
    \nonumber\\ & \phantom{=}
    + \mu \left( \frac{(1+p)^4}{(2+2p^2)^2} + \frac{(1-p)^4}{(2+2p^2)^2} \right)
    \nonumber\\ & \phantom{=}
   + \frac{t}{2} \frac{2 (1+p)(1-p)}{2+2p^2} \,.
\end{align}
Expanding in $p$, we get
\begin{align}
  \frac{E(p)}{N} &= e_0 + e_2 p^2 + e_4 p^4 + e_6 p^6 + \cdots \;,
 \end{align}
with
\begin{align}
  e_0 &= \frac{\mu}{2} -\frac{t}{2} + \frac{9V}{16} \;,
  \nonumber\\
  e_2 &= 2 \mu+t-\frac{V}{2} \;,
  \nonumber\\
  e_4 &= -4 \mu-t+2 V  \;,
    \nonumber\\
  e_6 &= 6 \mu+t-\frac{11 V}{2}\;.
\end{align}
The condition for a second-order phase transition is $e_2=0$ and  $e_4>0$. This happens along the
\begin{equation}
 \frac{V}{2} = 2 \mu+t
\end{equation}
line until the $e_4$ changes sign and becomes negative at $\mu= -3t/4$ and $V = -t$ (we note that $e_6$ is positive for $V<-t/2$). It signifies a tricritical point where the phase transition changes from a continuous (for $-t<V$) to a first-order one (for $V<-t$). Unfortunately, our ED calculation is unsuitable for confirming the tricritical point's existence beyond the variational approach.

\section{Details of perturbation of the classical phases}
\label{sec:perturb_appendix}

\begin{figure}[bt]
    \centering
    \includegraphics[width=.9\columnwidth]{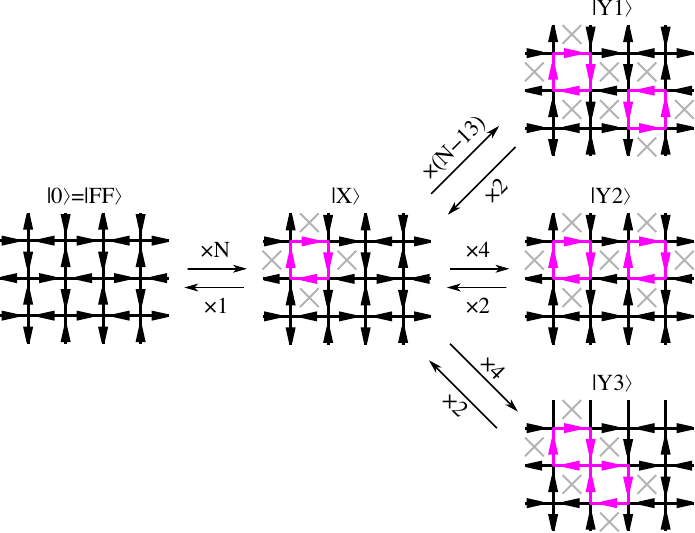}
    \caption{The excited states $\ket{X}$ and $\ket{Y}$ appearing in the fourth-order perturbation expansion Eq.~(\ref{eq:FourthOrdPert}) of the ground state energy in the fully flippable phase. Flipped plaquettes are in magenta and gray crosses indicate non-flippable plaquetes. The numbers on the long arrows show the number of transitions, where $N$ is the system size.
\label{fig:FF_Pert}}
\end{figure}

Perturbation theory is a convenient tool to determine energy corrections in the quasi-classical regime caused by quantum fluctuations. The ground state energy is expanded in powers of $t$ as 
\begin{equation}
   E  = \varepsilon^{(0)} + \varepsilon^{(2)} + \varepsilon^{(4)} + \cdots\;,
\end{equation}
where $\varepsilon^{(0)}$ is the classical energy and the second and fourth order corrections, following {\it e.g.} Ref.~\cite{Messiah1986}, read
\begin{align}
        \varepsilon^{(2)} & =  - \sum_{X} \dfrac{\left| \bra{0} \mathcal{H}_t \ket{X} \right|^2}{E_X - E_0}  \;,
    \label{eq:SecOrdPert} \\
    \varepsilon^{(4)} & =  - \sum_{X,Y,X'} \dfrac{\bra{0} \mathcal{H}_t \ket{X} \bra{X} \mathcal{H}_t \ket{Y} \bra{Y} \mathcal{H}_t \ket{X'} \bra{X'} \mathcal{H}_t \ket{0}}{(E_X - E_0)^2 (E_{Y} - E_0)}
    \nonumber \\ & \phantom{=}
    +  \left( \sum_{X} \dfrac{\left|\bra{0} \mathcal{H}_t \ket{X}\right|^2}{(E_X - E_0)^2}  \right)
    \left( \sum_{X} \dfrac{\left| \bra{0} \mathcal{H}_t \ket{X} \right|^2}{E_X - E_0}  \right).
    \label{eq:FourthOrdPert}
\end{align}
\label{eq:SecAndFourtPert}
Above,  we get $\ket{X}$ by flipping a single plaquette in the classical ground state configuration $\ket{0}$. The $\ket{Y} \neq \ket{0}$ are states constructed by flipping additional plaquettes in $\ket{X}$. Since the energy is an even function of $t$~\cite{Lan_PhysRevB.96.115140}, the correction terms proportional to an odd power in $t$ must vanish.

Let us start with fully the flippable states and choose $\ket{0} = |\text{FF1}\rangle$ in Eq.~(\ref{eq:ClassicalStates}). Fig.~\ref{fig:FF_Pert} shows the  possible $\ket{X}$ and $\ket{Y}$ states. The energies of these intermediate states are
\begin{subequations}
\begin{align}
E_X - E_0 & = 4\mu-4V \;, \\
E_{Y_1} - E_0 &= 8\mu-8V \;,\\
E_{Y_2} - E_0 &= 8\mu-7V \;,\\
E_{Y_3} - E_0 &= 6\mu-6V \;.
\end{align}
\end{subequations}
Using Eqs.~(\ref{eq:SecAndFourtPert}), the 2nd order correction to the energy is
\begin{equation}
    \varepsilon^{(2)}_{\rm FF}  =-\dfrac{Nt^2}{4(\mu - V)} \; , 
\end{equation}    
and the 4th order is    
\begin{widetext}
\begin{align}
        \varepsilon^{(4)}_{\rm FF} & =-\dfrac{N t^2}{16(\mu-V)^2}\left[ \dfrac{4\times2}{6(\mu-V)}t^2 + \dfrac{4\times2}{8\mu-7V}t^2 + \dfrac{(N-13)\times2}{8(\mu-V)}t^2 \right] 
         + \dfrac{N t^2}{16(\mu-V)^2} \times \dfrac{N t^2}{4(\mu-V)}
 \nonumber \\ &
 = -\dfrac{Nt^4}{16(\mu-V)^2} \left[ \dfrac{8}{8\mu - 7V} - \dfrac{23}{12(\mu-V)} \right] \;.
    \end{align}
\end{widetext}

\begin{figure}[t!]
    \centering
    \includegraphics[width=.9\columnwidth]{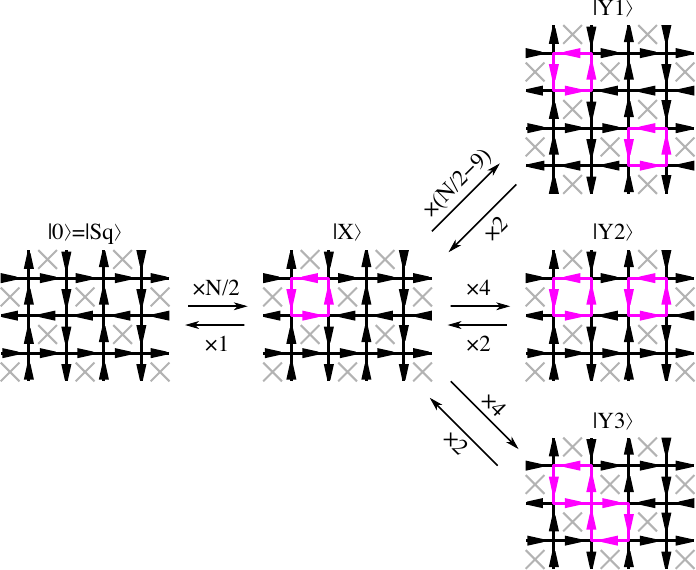}
    \caption{The same as Fig.~\ref{fig:FF_Pert}, but for the square phase. 
\label{fig:SQ_Pert}}
\end{figure}
 
 Let us turn to the square states. Choosing $\ket{0} = |\text{SqA}\rangle$ in Eq.~(\ref{eq:ClassicalStates}) and repeating the same steps for the square phase, we obtain the $\ket{X}$ and $\ket{Y}$ intermediate states shown in Fig.~\ref{fig:SQ_Pert},
with energies
\begin{subequations}
\begin{align}
E_X - E_0 & = -4\mu \;, \\
E_{Y_1} - E_0 &= -8\mu \;, \\
E_{Y_2} - E_0 &= -8\mu+V \;, \\
E_{Y_3} - E_0 &= -6\mu \;.
\end{align}
\end{subequations}
Since the square states are stable for negative $\mu$ values, all the intermediate energies above are positive. The 2nd and 4th-order corrections are then
\begin{subequations}
\begin{align}
   \varepsilon^{(2)}_{\rm Sq} & =
   \dfrac{Nt^2}{8\mu} \; ,  \\
   \varepsilon^{(4)}_{\rm Sq} & =
\dfrac{Nt^4}{16\mu^2} \left[ \dfrac{4}{8\mu - V} - \dfrac{11}{24\mu} \right] \;.
\end{align}
\end{subequations}

 \section{Symmetry properties of the $C(i_x,i_y)$ correlation dunctions}
\label{sec:app_hsym} 
 
The symmetry group $\mathsf{\tilde G}$ imposes the following properties to $C^{hh}$ defined in Eq.~(\ref{eq:defCxyhh}):
\begin{multline}
\label{eq:Chhsym}
 C^{hh}(i_x,i_y) = C^{hh}(-i_x,i_y)  
 = C^{hh}(i_x,-i_y) \\
  = C^{hh}(-i_x,-i_y) \;, 
 \end{multline}
which also holds for $C^{vv}$. Furthermore, the fourfold rotation symmetry connects the correlations between the vertical and the horizontal bonds
\begin{equation}
  \label{eq:ChhCvvsym}
   C^{vv}(i_x,i_y) =  C^{hh}(i_y,i_x).
\end{equation}
For the correlations between orthogonal bonds,
\begin{equation}
 C^{vh}(i_x,i_y) = C^{vh}(1-i_x,i_y) = C^{vh}(i_y+1,i_x-1) 
 \end{equation}
 holds, that generate eight positions with equal correlations, including the $C^{vh}(1-i_x,-1-i_y)$. Replacing the latter into the definition Eq.~(\ref{eq:defCxyvh}), we find that it is the same as Eq.~(\ref{eq:defCxyhv}) after shifting the coordinates, i.e.
 \begin{equation}
 C^{vh}(i_x,i_y) = C^{hv}(i_x,i_y).
   \label{eq:CvhChvsym}
 \end{equation}

\section{Analytic expressions for the structure factors}
\label{sec:SqPlaq}

\begin{figure}[bt!]
  \includegraphics[width = 0.9\columnwidth]{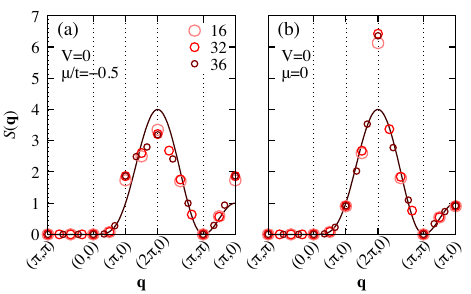}
  \caption{
 Comparison of the structure factor $S(\mathbf{q})$ in the plaquette phase calculated by ED (open circles for N=16, 32, and 36 sites) with the  $S(\mathbf{q})$ of the factorized plaquette wave function, Eq.~(\ref{eq:SqPl}) (solid line). We present two cases, (a) one close to the square phase with $\mu/t=-0.5$, and (b) one close to the flippable phase, where $\mu/t=0$ ($V=0$ for both).
  \label{fig:SQPlaq}
}
\end{figure}

In Sec.~\ref{sec:structure_factors}, we calculated the structure factors assuming that the ground state shows all the symmetries of the model and can be written as a linear superposition of the symmetry-broken states with equal weights. Here, we compute the structure factor for the symmetry-broken classical configurations and the plaquette states.

Translation invariance plays a crucial role when we calculate structure factors analytically. Therefore, we study the different ground states in translationally invariant unit cells containing $g_{\text{GS}}$ bonds representing the spins. For example, for the fully flippable configuration and the classical disordered phase boundary $g_{\text{FF}}=2$, since they do not break the translational symmetry, two bonds are associated with each vertex. The square phase breaks translational invariance, and there are four sites and eight bonds in the unit cell, so $g_{\text{Sq}}=8$. In the case of the plaquette phase, $g_{\text{Pl}}=4$. Rewriting the definition of the structure factor in Eq.~(\ref{eq:defSq}) to take the translationally invariant unit cell into account explicitly, we can derive that
\begin{align}
	S(\mathbf{q}) &= 
    \frac{1}{g_{\text{GS}}} \sum_{\mathbf{g}} \sum_{\mathbf{R}} 
    e^{i\mathbf{q}\cdot\mathbf{R}} 
    \langle n_{\mathbf{g} + \mathbf{R}} n_{\mathbf{g}} \rangle_\text{GS} 
   \nonumber\\ &\phantom{=}
    + \frac{1}{g_{\text{GS}}} \sum_{\mathbf{g}} \sum_{\mathbf{R}} 
    e^{i\mathbf{q}\cdot \left[\mathbf{R} + (\frac{1}{2},\frac{1}{2})\right]} 
    \langle n_{\mathbf{g} + \mathbf{R} + (\frac{1}{2},\frac{1}{2})} n_{\mathbf{g}} \rangle_\text{GS} \;,
\end{align}
in the fully packed loop representation,
where $\sum_\mathbf{g}$ denotes summation over the positions of spins within the invariant cell and $\sum_{\mathbf{R}}$ summation over the $N$ lattice vectors. Furthermore, we introduced the 
 $\langle \text{GS} | \hat{O} |  \text{GS} \rangle = \langle \hat{O} \rangle _{\text{GS}}$ short-hand notation, where $\text{GS}$ stands for one of the symmetry breaking ground state configurations. In the expression above, the first term comes from the correlation between parallel bonds, and the second term from perpendicular bonds. 

The case of magnetic structure factor $\tilde{S}(\mathbf{q})$ is slightly more complicated. First, let us write Eq.~(\ref{eq:FPL2Arrows}) as
\begin{equation}
	\mathbf{M}_{\mathbf{r} = \mathbf{R} + \mathbf{h}} = e^{i(\pi,\pi)\cdot \mathbf{R}}  
\left[ 
\begin{pmatrix}
1 \\ 
0
\end{pmatrix} 
\delta_{\mathbf{h},(\frac{1}{2},0)}	
	+ \begin{pmatrix}
0 \\ 
1
\end{pmatrix} 
\delta_{\mathbf{h},(0,\frac{1}{2})} \right]
n_{\mathbf{r}} ,
	\label{eq:Malpha}
\end{equation}
where $\mathbf{h} \in \left\{ (\frac{1}{2},0), (0,\frac{1}{2}) \right\}$. For $\tilde{S}(\mathbf{q})$, we obtain
\begin{multline}
\tilde{S}(\mathbf{q}) = \frac{1}{g_{\text{GS}}} \sum_{\mathbf{g}}\sum_{\mathbf{R}} \sum_{\alpha} 
  \left( 1 - \frac{q^2_{\alpha}}{q^2}\right) 
   e^{i\mathbf{q}\cdot\mathbf{R}} 
   \langle M^{\alpha}_{\mathbf{g} + \mathbf{R}} M^{\alpha}_{\mathbf{g}} \rangle_\text{GS} 
\\
-	\frac{1}{g_{\text{GS}}} \sum_{\mathbf{g}}\sum_{\mathbf{R}} \sum_{\alpha, \beta}   
	\frac{q_{\alpha} q_{\beta}}{q^2} 
	e^{i\mathbf{q}\cdot \left[\mathbf{R} + (\frac{1}{2},\frac{1}{2})\right]}
	\\
	\times \langle M^{\alpha}_{\mathbf{g} + \mathbf{R} + (\frac{1}{2},\frac{1}{2})} M^{\beta}_{\mathbf{g}} \rangle_\text{GS} 
	\; .
\end{multline}
Here again, the first term comes from the correlations between arrows on the parallel bonds ($\alpha=\beta$), and the second term from perpendicular bonds with orthogonal arrows ($\alpha \neq \beta$). 
Let us now evaluate the expressions above for the ordered phases.

\subsection{Fully flippable phase}
\label{sec:SQ_FF_calc}

 We start with the simplest case, the translationally invariant fully flippable state with two bonds in the unit cell, $\mathbf{g}\in \left\{ (\frac{1}{2},0), (0,\frac{1}{2}) \right\}$. For either of the classical states given by Eqs.~(\ref{eq:ClassicalStatesFF1})-(\ref{eq:ClassicalStatesFF2}), the correlation is $1$ between two horizontal or two vertical edges and $-1$ between a horizontal and a vertical edge. Since the structure factors of the two classical configurations are equal, it is enough to study only one.

In the fully packed loop representation, we get
\begin{align}
	S_{\text{FF}}(\mathbf{q}) 
	&= 2N\left( 1 - \cos \frac{q_x}{2} \cos \frac{q_y}{2} \right) \delta_{q_x,2\pi n_x} \delta_{q_y,2\pi n_y} 
	\nonumber\\
	&= 2N\left( 1 - (-1)^{(n_x + n_y)} \right) \delta_{q_x,2\pi n_x} \delta_{q_y,2\pi n_y} \;, \label{eq:SQ_FF_calc}
\end{align}
where $(n_x, n_y) \in \mathbb{Z}^2$. Only contributions from odd $n_x+n_y$ survive.   All the weight concentrates in a single peak (the other $\Gamma$ point) located at the corners of the extended Brillouin zone; see the third column in Fig.~\ref{fig:SQ}(b). For the $\tilde{S}_{\text{FF}}(\mathbf{q}) $, we obtain  
\begin{multline}
	\tilde{S}_{\text{FF}}(\mathbf{q}) 
	= \left( \frac{1}{2} + \frac{q_x q_y}{q^2} \sin \frac{q_x}{2} \sin \frac{q_y}{2} \right) N \\
	\times \delta_{q_x,2\pi n_x+\pi} \delta_{q_y,2\pi n_y+\pi} 
	\;.
\end{multline}
The fourfold symmetry is explicit in both $S_{\text{FF}}(\mathbf{q})$ and $\tilde{S}_{\text{FF}}(\mathbf{q})$. This might seem surprising, as the $|\text{FF1}\rangle$ and $|\text{FF2}\rangle$ themselves break the fourfold symmetry. But in the correlation function an even product of the $|\text{FF1}\rangle$ and $|\text{FF2}\rangle$ appears, restoring the $C_4$ symmetry. Since $\langle \text{FF1} | \text{FF2} \rangle = 0$, the structure factor for the linear superposition of the two configurations is the same.

\subsection{Square phase}
\label{sec:SQ_Sq_calc}

In the case of the square phase, Eqs.~(\ref{eq:ClassicalStatesSqA})-(\ref{eq:ClassicalStatesSqD}), the four orthogonal classical configurations are invariant under the translation by the $(2,0)$ and $(0,2)$ lattice vectors. The unit cell has eight sites ($g_{\text{Sq}} = 8$), with $ \mathbf{g} \in \left\{ (\frac{1}{2},0), (0,\frac{1}{2}), (\frac{1}{2}, 1), (1,\frac{1}{2}), (\frac{3}{2}, 0), (0,\frac{3}{2}), (\frac{3}{2}, 1), (1,\frac{3}{2}) \right\}$. We find
\begin{equation}
	S_{\text{Sq}}(\mathbf{q}) = \frac{N}{2}
	\left( 
	 \delta_{q_x,2\pi n_x} \delta_{q_y,2\pi n_y+ \pi}
	  + \delta_{q_x,2\pi n_x+\pi} \delta_{q_y, 2 \pi n_y} 
	 \right)
	 \label{eq:SQ_Sq_calc}
\end{equation}
and
\begin{align}
	\tilde{S}_{\text{Sq}}(\mathbf{q}) &= \frac{N}{2}
	\frac{q_x^2}{q^2} \delta_{q_x,\pi (2 n_x+ 1)} \delta_{q_y, 2\pi n_y} 
	\nonumber\\&\phantom{=}
	+
	\frac{N}{2}
    \frac{q_y^2}{q^2} \delta_{q_x,2\pi n_x} \delta_{q_y,\pi (2 n_y+ 1)} 
	\;.
\end{align}
for any of the four. The fourfold rotation symmetry manifests here again. The weight distributes equally in four peaks in the extended Brillouin zone, at $\mathbf{Q}= (\pm \pi,0)$ and $(0,\pm \pi)$, as seen in the leftmost column of Fig.~\ref{fig:SQ}(b).

\subsection{Plaquette phase}
\label{sec:SQ_Plaq_calc}

Next, let us consider the plaquette phase. The ground state is invariant under the translations (1,1) and (1,$-1$), with two sites and four bonds in the unit cell. Thus, $g_{\text{Pl}} = 4$, and $\{ \mathbf{g} \} = \left\{ (\frac{1}{2},0), (0,\frac{1}{2}), (-\frac{1}{2}, 0), (0,-\frac{1}{2}) \right\}$ . We use the variational wave functions $| \text{PlAD} \rangle$ and $| \text{PlBC} \rangle$, defined in 
 Eqs.~(\ref{eq:Plaq_WFAD})-(\ref{eq:Plaq_WFBC}). Here one should be careful when taking their linear combination, as they have a finite overlap $\langle \text{PlAD} | \text{PlBC} \rangle = 4/N^2$. The structure factors for the pure $|\text{PlAD}\rangle$ are
\begin{align}
	S_{\text{PlAD}}(\mathbf{q}) &=  4 \sin^2 \frac{q_x + q_y}{4} \sin^2 \frac{q_x - q_y}{4}, \label{eq:SqPl}\\
	\tilde{S}_{\text{PlAD}}(\mathbf{q}) 
	&= \left( \frac{q_x}{q} \sin \frac{q_x}{2} + \frac{q_y}{q} \sin \frac{q_y}{2}  \right)^2,
\end{align}
and $S_{\text{PlAD}}(\mathbf{q}) = S_{\text{PlBC}}(\mathbf{q})$, $\tilde{S}_{\text{PlAD}}(\mathbf{q}) = \tilde{S}_{\text{PlBC}}(\mathbf{q})$. For a general plaquette state, we can expand them as a superposition of $| \text{PlAD} \rangle$ and $| \text{PlBC} \rangle$. 
Since the terms coming from the overlap of the two  wave functions are proportional to $1/N^2$ and vanish in the thermodynamic limit,  $S_{\text{Pl}}^{\infty}(\mathbf{q}) = S_{\text{PlAD}}(\mathbf{q})$ and
	$\tilde{S}_{\text{Pl}}^{\infty}(\mathbf{q}) = \tilde{S}_{\text{PlAD}}(\mathbf{q})$. We compare in Fig.~\ref{fig:SQPlaq} the $S_{\text{Pl}}^{\infty}(\mathbf{q})$ with the result from exact diagonalization. Eq.~(\ref{eq:SqPl}) nicely captures the main features of the diffuse scattering in the structure factor.

\subsection{Disordered manifold at $t=V=0$}
\label{sec:SQ_Dirord_calc}

Finally, let us determine the structure factor of disordered type-II vertices at the $V=0$ boundary between the isolated and square phases in the classical phase diagram. 
Two bonds are correlated only if they belong to the same horizontal or vertical line.  
\begin{align}
	S_{\text{Dis}}(\mathbf{q}) &= \frac{1}{2} \left( L_x \delta_{q_x, 2\pi z} + L_y \delta_{q_y, 2\pi z} \right), \label{eq:SqDis}\\
	\tilde{S}_{\text{Dis}}(\mathbf{q}) 
	&= \frac{1}{2} \left[ \frac{q_y^2}{q^2} \times L_x \delta_{q_x, 2\pi z_x} + \frac{q_x^2}{q^2} \times L_y \delta_{q_y, 2\pi z_y} \right],
\end{align}
where $L_x$ and $L_y$ are the horizontal and vertical lengths of the periodic cluster, and $z_x, z_y \in \mathbb{Z}$ (we set $g_{\rm Dis} = 2$ since the disordered manifold is translational invariant as a set of ice configurations).

All the structure factors we computed above reflect all the symmetries of $D_4$ point group, just like for the $S(\mathbf{q})$ and $\tilde{S}(\mathbf{q})$ calculated from the fully symmetric ground state of a finite cluster with periodic boundary conditions. Furthermore, all of the $S(\mathbf{q})$ above satisfy the sum rule given by Eq.~(\ref{eq:sumrule}).

\bibliography{bibliography_v2}
\end{document}